\pdfoutput=1
\documentclass[11pt]{article}

\usepackage[margin=1in]{geometry}
\usepackage[T1]{fontenc}
\usepackage{lmodern}
\usepackage{microtype}
\usepackage{amsmath,amssymb,amsthm,mathtools,bm}
\usepackage{booktabs,array,threeparttable}
\usepackage{graphicx}
\usepackage{xcolor}
\usepackage{tikz}
\usetikzlibrary{calc}
\usepackage{pgfplots}
\pgfplotsset{compat=1.18}
\usepackage{caption}
\usepackage{subcaption}
\usepackage{enumitem}
\usepackage[authoryear,round]{natbib}
\usepackage[hidelinks]{hyperref}
\usepackage[capitalise,noabbrev]{cleveref}
\usepackage{setspace}
\theoremstyle{plain}
\newtheorem{theorem}{Theorem}
\newtheorem{proposition}{Proposition}
\newtheorem{lemma}{Lemma}
\newtheorem{corollary}{Corollary}
\newtheorem{conjecture}{Conjecture}
\theoremstyle{definition}
\newtheorem{definition}{Definition}
\newtheorem{numresult}{Numerical result}
\theoremstyle{remark}
\newtheorem{remark}{Remark}
\crefname{numresult}{Numerical result}{Numerical results}
\crefname{conjecture}{Conjecture}{Conjectures}

\newcommand{\kstar}{\kappa^{*}}
\newcommand{\one}{\mathbf{1}}
\newcommand{\E}{\mathbb{E}}
\DeclareMathOperator{\Var}{Var}
\DeclareMathOperator{\SD}{SD}
\newcommand{\tauhat}{\hat\tau_{\mathrm{ETI}}}
\newcommand{\betahat}{\hat\beta}
\newcommand{\ebar}{\bar e}
\newcommand{\Wtw}{W^{\mathrm{W}}}
\newcommand{\Wbt}{W^{\mathrm{B}}}

\title{Adequacy radii for stepped-wedge trials with time-varying effects}
\author{Stanis{\l}aw M. S. Halkiewicz\thanks{Jagiellonian University, Krak\'ow, Poland.}
\and Sergey Alexeev\thanks{University of New South Wales, Sydney, and NHMRC Clinical Trials Centre, The University of Sydney, Australia. Correspondence: sergei.v.alexeev@gmail.com.}}
\date{}

\begin{document}
\maketitle

\begin{abstract}
Complete stepped-wedge designs guarantee eventual intervention delivery, but their final periods contain no concurrent controls. Under exposure-time heterogeneity, this creates a design-analysis tension: a constant-effect analysis can be precise while targeting a design-dependent contrast, whereas an exposure-time indicator (ETI) analysis can target a prespecified time-averaged effect but may be substantially less precise. We develop an adequacy radius that quantifies the size of exposure-time heterogeneity compatible with a prespecified tolerance for estimand mismatch. The radius makes the loss of concurrent controls visible and can be combined with precision calculations to compare complete wedges with rollout-constrained hybrid designs that retain a wait-list arm. For a balanced wedge with a wait-list under the Hussey--Hughes model, we show that the wait-list share and the correlation act on the constant-effect GLS estimand through a single scalar, and prove that the radius increases strictly with the wait-list share and decreases strictly with the within-cluster correlation; within this family it remains bounded, so the wait-list makes the constant-effect analysis more tolerant but not robust. Numerical optimisation and a model-based redesign of the Haines trial show how intermediate hybrids can make estimand-aligned ETI analysis materially more affordable, producing a robustness--precision frontier rather than a binary choice between a complete wedge and a parallel trial. A second trial, of digital adherence support for tuberculosis treatment, has the same radius but no detectable heterogeneity; its test of a constant effect would have detected heterogeneity large enough to halve the coverage of the nominal interval only 37\% of the time, so a null check is uninformative in a complete design. The contribution is not a new model for exposure-time heterogeneity, a new ETI estimator, or the invention of hybrid designs. It is a diagnostic and design principle: when exposure-time heterogeneity is scientifically plausible, retaining concurrent controls can reduce the price of analysing the estimand one actually intends to report.
\end{abstract}


\clearpage

\section{Introduction}\label{sec:intro}

In a stepped-wedge cluster randomised trial, clusters cross from the control to the intervention condition at randomly assigned times, and by the end of the trial every cluster has been treated \citep{hussey2007design,hemming2015stepped}. The standard analysis is a linear mixed model with period effects and a single treatment indicator \citep{hussey2007design,li2021mixed}. This immediate-treatment (IT) model assumes that the full effect of the intervention appears in the first period after crossover and does not change afterwards. Many interventions evaluated in stepped wedges are service or behaviour changes whose effect builds up, or wears off, with time since implementation, that is, with exposure time \citep{hughes2015current}. \citet{kenny2022analysis} showed that under exposure-time heterogeneity the expectation of the IT estimator is a weighted sum of the exposure-specific effects whose weights sum to one but can be negative, so the estimate can lie far from the time-averaged treatment effect (TATE) and can have the opposite sign. \citet{lee2025analysis} distinguished heterogeneity in exposure time from heterogeneity in calendar time. The same phenomenon has been studied for two-way fixed-effects regressions in staggered-adoption designs \citep{goodmanbacon2021difference,dechaisemartin2020two,sun2021estimating,callaway2021difference,borusyak2024revisiting}, and \citet{lindner2021heterogeneous} made the link to the stepped wedge explicit.

Remedies at the analysis stage are available. The exposure-time-indicator (ETI) model of \citet{kenny2022analysis} estimates a separate effect for each exposure time and averages them; random-effect and spline formulations pool information across exposure times \citep{maleyeff2023assessing,kenny2022analysis}; model-robust and semiparametric estimators target exposure-time estimands without relying on the working covariance \citep{wang2024achieve,fang2026model,xia2025robust}; a generalised difference-in-differences estimator targets chosen estimands directly \citep{kennedyshaffer2026generalized}; and small-sample corrections for ETI models have been compared \citep{ouyang2026which}. The difficulty is precision. \citet{kenny2026factors} found that keeping 90\% power when moving from an IT model to an ETI model targeting TATE commonly requires 1.5 to 3 times the sample size and often more; that additional baseline periods help; that additional periods after all clusters are treated help little; and that optimal designs for time-varying estimands remain to be determined. Design choice under exposure-time heterogeneity has itself been studied. \citet{lee2024cluster} derived the variances of time-on-intervention (exposure-time) effect estimators in four fixed designs: a stepped wedge, a parallel design, a parallel design with a baseline period, and a parallel design with a baseline and a final period in which all clusters are treated. They showed that the stepped wedge estimates the first exposure effect most efficiently, while the parallel design with baseline usually estimates the long-term and time-averaged effects most efficiently, and that in parallel designs the constant-effect and time-averaged estimators coincide. They illustrated the comparison with simulations calibrated to the weekend allied-health disinvestment trial that we also use in \cref{sec:example}.

A separate literature optimises the design of longitudinal cluster trials for precision under a constant treatment effect. \citet{girling2016statistical} showed that the precision-optimal design in large studies is a hybrid of stepped-wedge and parallel components, with the stepped-wedge share equal to the cluster-mean correlation; \citet{thompson2017optimal} found that the most efficient stepped wedge with equal allocation keeps the first sequence in the control condition and the last in the intervention condition throughout; \citet{watson2023optimal} review algorithms for $c$-optimal and robust designs, where robustness refers to uncertainty in the covariance parameters; and \citet{girling2025optimal}, \citet{liu2025selecting} and \citet{xiong2024optimal} optimise observation allocation, design family and rollout timing for precision. These optimisation results assume a constant effect. The design work that allows exposure-time heterogeneity \citep{lee2024cluster,kenny2026factors} compares the precision of correctly specified exposure-time estimators across a small number of given designs.

This paper also treats exposure-time heterogeneity as a design problem, but asks two questions that those comparisons do not: how much heterogeneity a given design and analysis can tolerate before the nominal inference of the conventional IT analysis fails, and which features of the design change that amount and the cost of the robust alternative. The measuring device is the \emph{adequacy radius}: the largest root-mean-square (RMS) variation of the exposure-time effects, as a fraction of the effect the trial is powered to detect, for which the nominal 95\% confidence interval still covers TATE at least 90\% of the time. It specialises the design diagnostic of \citet{halkiewicz2026positive} to exposure-time heterogeneity and to planned power, and we define it here without relying on that paper. Our results are as follows.

\begin{enumerate}[label=(\roman*),leftmargin=2em]
\item \emph{Complete stepped wedges tolerate little heterogeneity, and almost the same amount whatever their size.} For balanced complete designs we obtain the radius in closed form. Under two-way fixed effects (TWFE) it lies in $(0.1194, 0.1235]$ for every number of periods $T\ge 4$; under the Hussey--Hughes model it is below $0.2604$ and strictly decreasing in the intracluster correlation. Under a linear learning curve the TWFE-IT estimator equals the effect curve evaluated at exposure $1/2$ for every $T$, and among effect curves with bounded increments the linear ramp is the worst case. Numerically, reallocating clusters across steps changes the radius little.
\item \emph{The mechanism is the loss of concurrent controls.} Every linear estimator that is invariant to period effects gives zero net weight to any period in which all clusters are treated. In a complete design the final period therefore identifies only differences between exposure-time effects, and the longest exposures enter every estimator only through such differences. Without a restriction on the form of heterogeneity, effects in the final period are not identified at all, which is the stepped-wedge instance of a result of \citet{borusyak2024revisiting}.
\item \emph{Concurrent controls change the geometry.} With a parallel layout the IT weights are uniform for every correlation \citep[an equivalent statement is in][]{lee2024cluster}. For a balanced wedge with a wait-listed share $q$ we give the IT weights and radius in closed form, under TWFE and under the Hussey--Hughes model. Under TWFE the weights are a mixture of the complete-wedge weights and a kernel centred on the target. Under the Hussey--Hughes model all designs in the family and all correlations give weights on a single one-parameter curve, and the radius increases strictly with $q$ and decreases strictly with the correlation, but stays bounded.
\item \emph{Concurrent controls make the robust analysis cheaper.} Numerically, at the rollout capacity of a balanced wedge with 24 clusters and five periods, wait-listing 4 clusters reduces the variance cost of ETI analysis from 2.27 to 1.52 times the IT variance of the complete wedge, and the designs that minimise ETI variance wait-list 42--54\% of clusters at a cost of 1.02--1.41 across correlation structures. These designs differ materially from the constant-effect optimum.
\item \emph{Two applications.} In a trial that withdrew weekend allied-health services from 12 wards \citep{haines2017impact}, we reproduce the published IT and ETI analyses, compute the radius of the trial's actual design ($0.124$), and show that the estimated exposure-time variation exceeds it about fivefold. A counterfactual exercise calibrated to the trial, with outcomes under the alternative designs simulated rather than observed, shows that wait-listing two to five of the 12 wards at the same rollout capacity would have reduced the variance of the ETI estimator by 30--42\%. In a trial of digital adherence support for tuberculosis treatment in 18 facilities \citep{cattamanchi2021digital}, with a binary outcome, we reproduce the published result and find the same radius ($0.122$) but no detectable heterogeneity. The trial could not have detected heterogeneity that mattered: against the least-favourable profile at three times the radius, which would halve the coverage of the nominal IT interval, the test of a constant effect has power $37\%$. In that trial a wait-list of four to eight facilities would have cut the ETI variance by 44--54\% without increasing the IT variance.
\end{enumerate}

Several ingredients are not new, and we cite them where they are used: exposure-time heterogeneity in stepped wedges and the weighting representation of the IT estimator \citep{kenny2022analysis}, including the closed-form weights for the balanced design; ETI analysis and its cost \citep{kenny2022analysis,kenny2026factors}; comparison of designs for exposure-time effects, the advantage of parallel designs with baseline for time-averaged effects, the equivalence of IT and time-averaged estimators in parallel designs, and a simulation calibrated to the trial of \cref{sec:example} \citep{lee2024cluster}; hybrid stepped-wedge/parallel designs and their precision \citep{girling2016statistical,thompson2017optimal}; and the identification problem without never-treated units \citep{borusyak2024revisiting}. Relative to \citet{lee2024cluster}, whose criterion is the variance of correctly specified exposure-time estimators in four given designs, we add three things. The first is a criterion for the misspecified IT analysis, the adequacy radius, with closed forms and bounds for complete and hybrid designs. The second is a structural result, the zero net weight of all-treated periods for every period-invariant estimator, that explains why complete designs score low on that criterion and why their robust analysis is expensive. The third is optimisation over intermediate allocations of clusters to start periods and a wait-list, under the rollout capacity that motivates a stepped wedge, which locates the front-loaded hybrids between the stepped wedge and the parallel design with baseline.

Results labelled theorem, proposition, lemma or corollary are proved in \cref{app:proofs}; their closed forms have also been checked in exact rational arithmetic for $T=3,\dots,10$. Results labelled ``numerical result'' are computed, and conjectures are stated as such. Section~\ref{sec:setup} sets up designs, model and estimands; \cref{sec:radius} defines the radius; \cref{sec:complete,sec:hybrid} treat complete and hybrid designs; \cref{sec:cost} studies the cost of robust analysis; \cref{sec:example,sec:uganda} apply the results to two trials; \cref{sec:discussion} discusses implications and limitations.

\paragraph{Organising contribution.}
The paper's claim is deliberately narrower than several nearby literatures. Exposure-time heterogeneity, unequal or negative implicit weights, ETI/TATE analyses, hybrid stepped-wedge/parallel designs, and generic rollout optimisation are all established ideas. Our contribution is to connect three elements that are usually treated separately: a calibrated adequacy diagnostic for the constant-effect estimand; a structural explanation centred on the disappearance of concurrent controls; and optimisation over intermediate, rollout-constrained hybrid designs using the precision cost of an estimand-aligned ETI analysis. \Cref{thm:hybridgls} proves that, in the balanced compound-symmetry family, the radius against TATE increases with the wait-list share and decreases with the correlation; Appendix~\ref{app:balanced-waitlist-monotonicity} gives the corresponding result for the distance to the design-average exposure target. Broader claims about ETI precision and globally optimal wait-list allocation remain computational and are labelled as such.

\section{Designs, model and estimands}\label{sec:setup}

\subsection{Designs}

A trial has $G$ clusters observed in periods $t=1,\dots,T$. Each cluster belongs to a sequence $g\in\mathcal{S}=\{2,\dots,T\}\cup\{\infty\}$: a cluster in sequence $g$ is in the control condition in periods $t<g$ and in the intervention condition from period $g$ onwards. Sequence $g=\infty$ is a \emph{wait-list}: its clusters remain in the control condition throughout follow-up and may receive the intervention after the trial. The treatment indicator is $D_{gt}=\one\{t\ge g\}$, and a treated cell $(g,t)$ has exposure time $e=t-g+1$. A design is a vector of shares $p=(p_g)_{g\in\mathcal{S}}$ with $\sum_g p_g=1$; sequence $g$ contains $n_g=Gp_g$ clusters. We assume $p_2>0$, so the longest exposure $L=T-1$ is observed, and that every cluster is observed in every period with the same number $m$ of individuals per cluster-period (cross-sectional sampling). The theory treats $p$ as continuous, in the sense of approximate design theory; the numerical results use integer $n_g$.

A design is \emph{complete} if $p_\infty=0$. Three families recur:
\begin{itemize}[leftmargin=1.5em]
\item the balanced complete stepped wedge $\mathcal{W}_T$: $p_g=1/(T-1)$ for $g=2,\dots,T$;
\item the hybrid family $\mathcal{H}_T(q)$: $p_g=(1-q)/(T-1)$ for $g=2,\dots,T$ and $p_\infty=q$, $0\le q<1$;
\item the parallel design with baseline $\mathcal{P}_T(p)$: $p_2=p$ and $p_\infty=1-p$.
\end{itemize}
In this paper ``concurrent control'' means a control cluster observed in the same period as a treated cluster. In a complete design the final period has none. \Cref{fig:schematic} shows $\mathcal{W}_5$ and a hybrid.

\begin{figure}[t]
\centering
\newcommand{\swgrid}[3]{%
\begin{tikzpicture}[x=0.85cm,y=0.5cm,font=\scriptsize]
  \node[anchor=south] at (2.5,1.1) {#3};
  \foreach \t in {1,...,5} {\node at (\t-0.5,0.5) {$t=\t$};}
  \foreach \g [count=\r] in {#1} {
    \foreach \t in {1,...,5} {
      \pgfmathtruncatemacro{\e}{\t-\g+1}
      \ifnum\e>0
        \fill[gray!45] (\t-1,-\r) rectangle (\t,-\r+1);
        \node at (\t-0.5,-\r+0.5) {\e};
      \else
        \node[gray] at (\t-0.5,-\r+0.5) {0};
      \fi
      \draw[gray!70] (\t-1,-\r) rectangle (\t,-\r+1);
    }
    \node[anchor=east] at (0,-\r+0.5) {$g=\g$};
  }
  \ifnum#2>0
  \foreach \w in {1,...,#2} {
    \pgfmathtruncatemacro{\rr}{4+\w}
    \foreach \t in {1,...,5} {
      \node[gray] at (\t-0.5,-\rr+0.5) {0};
      \draw[gray!70] (\t-1,-\rr) rectangle (\t,-\rr+1);
    }
    \node[anchor=east] at (0,-\rr+0.5) {$g=\infty$};
  }
  \fi
  \draw[red!70!black, thick, dashed] (4,-4-#2) rectangle (5,0);
\end{tikzpicture}}
\begin{subfigure}[b]{0.45\textwidth}\centering
\swgrid{2,3,4,5}{0}{(a) Complete stepped wedge $\mathcal{W}_5$}
\end{subfigure}\hfill
\begin{subfigure}[b]{0.45\textwidth}\centering
\swgrid{2,3,4,5}{2}{(b) Stepped wedge with wait-list}
\end{subfigure}
\caption{Two designs with $T=5$ periods. Rows are sequences (one row may hold several clusters); shaded cells are treated and show the exposure time $e$. The final period is outlined. In (a) every cluster is treated in period 5, so that period contains no concurrent control and the longest exposure ($e=4$) is observed only there. In (b) wait-listed clusters ($g=\infty$) supply concurrent controls in every period; they can receive the intervention after follow-up.}
\label{fig:schematic}
\end{figure}

\subsection{Model and estimand}

Let $\bar Y_{ct}$ be the mean outcome of the $m$ individuals observed in cluster $c$ in period $t$, and let $g(c)$ be the sequence of cluster $c$. We assume
\begin{equation}\label{eq:model}
\bar Y_{ct}=\mu_t+\alpha_c+D_{g(c)t}\,\theta_{t-g(c)+1}+\varepsilon_{ct},
\end{equation}
where $\mu_t$ are period effects, $\varepsilon_{ct}$ are independent with mean zero and variance $\sigma^2$ (set to 1 without loss of generality), and $\theta=(\theta_1,\dots,\theta_L)^\top$ are the exposure-time effects. The cluster effects $\alpha_c$ are either independent with mean zero and variance $\phi\sigma^2$, giving the exchangeable within-cluster covariance $V=I_T+\phi J_T$ of \citet{hussey2007design}, or fixed. With individual-level intracluster correlation $\rho$, $\phi=m\rho/(1-\rho)$. The correlation between two cluster-period means of the same cluster is $\psi=\phi/(1+\phi)$, which is the parameter written $\phi$ by \citet{kenny2022analysis}; the cluster-mean correlation of \citet{girling2016statistical} is $R=T\phi/(1+T\phi)$. We write $\lambda=1/(1+T\phi)=1-R$.

Model~\eqref{eq:model} allows the effect to vary with exposure time but not with calendar time or between clusters; we return to calendar-time heterogeneity in \cref{prop:identification} and \cref{sec:discussion}. The target is the time-averaged treatment effect over the observed exposure range,
\[
\tau(\theta)=\frac1L\sum_{e=1}^L\theta_e .
\]
This is the discrete form of the TATE of \citet{kenny2022analysis} over the interval $(0,L]$ and coincides with it when the effect curve is constant within periods.

\subsection{Analyses and exposure weights}

The IT analysis fits \eqref{eq:model} with $\theta_e\equiv\beta$; it is fitted by generalised least squares with working parameter $\phi$ (IT-GLS) or with fixed cluster effects (TWFE-IT, the limit $\phi\to\infty$). The ETI analysis fits \eqref{eq:model} with unrestricted $\theta$ and reports $\tauhat=L^{-1}\sum_e\hat\theta_e$ \citep{kenny2022analysis}.

All these estimators are linear, $\betahat=\sum_c\sum_t a_{ct}\bar Y_{ct}$. Call $\betahat$ \emph{period-invariant} if its expectation does not depend on $(\mu_t)$, that is, $\sum_c a_{ct}=0$ for every $t$. (With fixed cluster effects we also need $\sum_t a_{ct}=0$ for every $c$.) The \emph{exposure weights} of $\betahat$ are
\[
W_e=\sum_{(c,t):\,D_{g(c)t}=1,\ t-g(c)+1=e} a_{ct},\qquad e=1,\dots,L,
\]
and for a period-invariant estimator \eqref{eq:model} gives $\E\betahat=\sum_eW_e\theta_e=W^\top\theta$. An estimator is unbiased under a constant effect if and only if $\one^\top W=1$, and unbiased for $\tau$ under every $\theta$ if and only if $W=u:=L^{-1}\one$. The IT estimators satisfy the first condition and the ETI estimator the second. By symmetry, GLS and TWFE give equal weights to all clusters of a sequence, so the exposure weights depend on the design only through $p$.

\section{The adequacy radius}\label{sec:radius}

Measure the size of exposure-time heterogeneity by the equal-weight RMS deviation of $\theta$ from its mean,
\[
s(\theta)=\Big\{\frac1L\sum_{e=1}^L(\theta_e-\tau)^2\Big\}^{1/2},\qquad \Theta(s)=\{\theta: s(\theta)\le s\}.
\]

\begin{proposition}[Worst-case bias]\label{prop:worstbias}
Let $\betahat$ be period-invariant with $\one^\top W=1$, and let $\Gamma=\sqrt L\,\lVert W-u\rVert_2$. Then
\[
\sup_{\theta\in\Theta(s)}\big|\E\betahat-\tau(\theta)\big|=\Gamma s,
\]
attained by $\theta-\tau\one=\pm s\sqrt L\,(W-u)/\lVert W-u\rVert_2$.
\end{proposition}

The quantity $\Gamma$ depends only on the design, the analysis model and the target. It is the exposure-time version of the design statistic of \citet{halkiewicz2026positive}. To turn bias into a statement about inference, consider the nominal two-sided level-$\alpha$ Wald interval $\betahat\pm z_{1-\alpha/2}\SD(\betahat)$. If its bias is $b$, it covers $\tau$ with probability $c(b/\SD(\betahat))$, where $c(\eta)=\Phi(z_{1-\alpha/2}-\eta)-\Phi(-z_{1-\alpha/2}-\eta)$ is even and decreasing in $|\eta|$. For a tolerated loss of coverage $\epsilon$, let $\eta_{\alpha,\epsilon}$ solve $c(\eta)=1-\alpha-\epsilon$.

\begin{definition}[Adequacy radius]\label{def:radius}
The adequacy radius of a design and analysis is
\[
r^\dagger=\frac{\eta_{\alpha,\epsilon}\,\SD(\betahat)}{\Gamma}\qquad(r^\dagger=\infty\text{ if }\Gamma=0),
\]
the largest $s$ for which the nominal interval covers $\tau$ with probability at least $1-\alpha-\epsilon$ for every $\theta\in\Theta(s)$. If the trial is planned to have power $1-\beta$ for a constant effect $\delta$, then $\SD(\betahat)=\delta/(z_{1-\alpha/2}+z_{1-\beta})$, and the \emph{relative radius at planned power} is
\begin{equation}\label{eq:kstar}
\kstar=\frac{r^\dagger}{\delta}=\frac{\eta_{\alpha,\epsilon}}{(z_{1-\alpha/2}+z_{1-\beta})\,\Gamma}.
\end{equation}
\end{definition}

We use $\alpha=0.05$, $\epsilon=0.05$ and $1-\beta=0.8$ throughout. Then $\eta_{\alpha,\epsilon}=0.6524$, $z_{0.975}+z_{0.8}=2.8016$ and $\kstar=0.2329/\Gamma$. %
A radius $\kstar=0.12$ means that exposure-time effects whose RMS deviation from TATE equals 12\% of the planned effect can already reduce the coverage of the nominal 95\% interval to 90\%.

\begin{remark}[What the radius depends on]\label{rem:radius}
(i) Through \eqref{eq:kstar}, $\kstar$ depends on the design only through $\Gamma$. It does not depend on $G$, $m$ or $\sigma^2$ once the trial is powered for $\delta$; in particular it is unchanged by adding clusters to every sequence in proportion. The absolute radius $r^\dagger=\kstar\delta$ shrinks as $\SD(\betahat)$ does: a larger trial converts the same heterogeneity into more standard errors of bias. (ii) The radius is a design-stage breakdown value for the nominal interval. It is related to bias-aware (``honest'') intervals \citep{armstrong2018optimal,armstrong2020simple}, which widen the interval by the worst-case bias over a declared class, and to sensitivity analyses that report the size of violation at which a conclusion changes \citep{rambachan2023more}. We use both connections in \cref{sec:cost}. (iii) The calculation uses the true standard deviation of $\betahat$. Under heterogeneity, model-based variance estimates from the IT fit are themselves misspecified \citep{kenny2022analysis}; \cref{sec:smallG} checks the effect of cluster-robust standard errors. (iv) $\Theta(s)$ allows any profile, including non-monotone ones. \Cref{prop:ramp} shows that for complete wedges the worst case over smooth profiles is the linear learning curve and that, under TWFE, it attains 89--100\% of the worst case over $\Theta(s)$.
\end{remark}

\section{Complete designs}\label{sec:complete}

\subsection{All-treated periods carry no level information}

\begin{lemma}[Period balance]\label{lem:balance}
For every period-invariant linear estimator, $\sum_c a_{ct}=0$ for each period $t$. Consequently, if every cluster is treated in period $t$, the treated cells of period $t$ carry zero net weight: $\sum_c a_{ct}D_{g(c)t}=0$.
\end{lemma}

\begin{corollary}[Final period of a complete design]\label{cor:final}
In any complete design, for every period-invariant estimator (IT or ETI, GLS under any working covariance, or TWFE), the exposure weights contributed by period $T$ sum to zero. The longest exposure $L$ is observed only in cell $(2,T)$, so its weight $W_L=n_2a_{2T}$ is offset by weight $-W_L$ on shorter exposures in the same period. The same holds for any additional period appended after period $T$.
\end{corollary}

\Cref{lem:balance} is elementary, but it organises most of what follows. A period in which all clusters are treated contains no contrast between treated and untreated clusters; once period effects are removed it can only compare clusters at different exposure times.

\begin{proposition}[What all-treated periods identify]\label{prop:identification}
(a) Under \eqref{eq:model}, in a period $t$ in which all clusters are treated, conditional on the cluster effects, $\E(\bar Y_{ct}-\bar Y_{c't})=\alpha_c-\alpha_{c'}+\theta_{t-g(c)+1}-\theta_{t-g(c')+1}$. The period identifies only differences between exposure-time effects.

(b) Replace $D_{gt}\theta_{t-g+1}$ in \eqref{eq:model} by unrestricted cohort-period effects $D_{gt}\tau_{gt}$. In a complete design, the final-period effects $(\tau_{gT})_{g}$ are identified only up to a common additive constant, so no $\tau_{gT}$ and no weighted average of them with positive weights is identified. If $p_\infty>0$, every $\tau_{gt}$ with $2\le g\le t\le T$ is identified by the concurrent contrast $\E\{(\bar Y_{gt}-\bar Y_{g1})-(\bar Y_{\infty t}-\bar Y_{\infty1})\}$, where $\bar Y_{gt}$ denotes the mean over clusters of sequence $g$.
\end{proposition}

Part (b) is the stepped-wedge instance of the under-identification of effects in staggered-adoption designs without never-treated units \citep{borusyak2024revisiting};
\citet{lee2025analysis} note the same for the final-period calendar-time effect and mention an unexposed cluster as a remedy. Part (a) explains how the ETI model nonetheless identifies $\theta_L$ in a complete design: the final period gives $\theta_L-\theta_{e}$ for shorter exposures $e$, and those are tied to concurrent contrasts in earlier periods. The level of the longest exposure effect is carried through this chain and depends on the restriction that effects vary with exposure time only. The two-sequence calculation of \citet[Appendix~A]{kenny2026factors} shows the chain explicitly, and their finding that periods added after all clusters are treated add little power for TATE is a consequence of \cref{cor:final}. For the IT estimator, which assumes all exposure effects equal, the final period is used to estimate period and cluster effects; the longest exposure receives zero or negative weight, and under TWFE every exposure beyond about half the trial is weighted negatively (\cref{thm:twfe,thm:gls}).

\subsection{The balanced wedge under two-way fixed effects}

\begin{theorem}[TWFE, balanced complete wedge]\label{thm:twfe}
For $\mathcal{W}_T$ with $T\ge3$, the TWFE-IT exposure weights are
\begin{equation}\label{eq:Wtwfe}
\Wtw_e=\frac{6(T-e)\{(T^2-1)-(2T-1)e\}}{T(T+1)(T-1)(T-2)},\qquad e=1,\dots,T-1,
\end{equation}
and
\begin{equation}\label{eq:G2twfe}
\Gamma^2=\frac{19T^3-31T^2-2T+12}{5T(T-2)(T+1)}.
\end{equation}
In particular $\Wtw_1=6/(T+1)$, $\Wtw_{L}=-6/\{T(T+1)\}$, and $\Wtw_e<0$ exactly when $e>(T^2-1)/(2T-1)$, that is, for exposures beyond about half the trial. For $T\ge4$, $19/5-\Gamma^2=12(T^2-3T-1)/\{5T(T-2)(T+1)\}>0$ and $\Gamma^2\ge 249/70$, with equality at $T=6$. Hence
\[
0.1194<\kstar\le 0.1235\qquad\text{for every }T\ge4,
\]
and $\kstar=0.1164$ at $T=3$. The weights and $\kstar$ do not depend on the number of clusters per sequence.
\end{theorem}

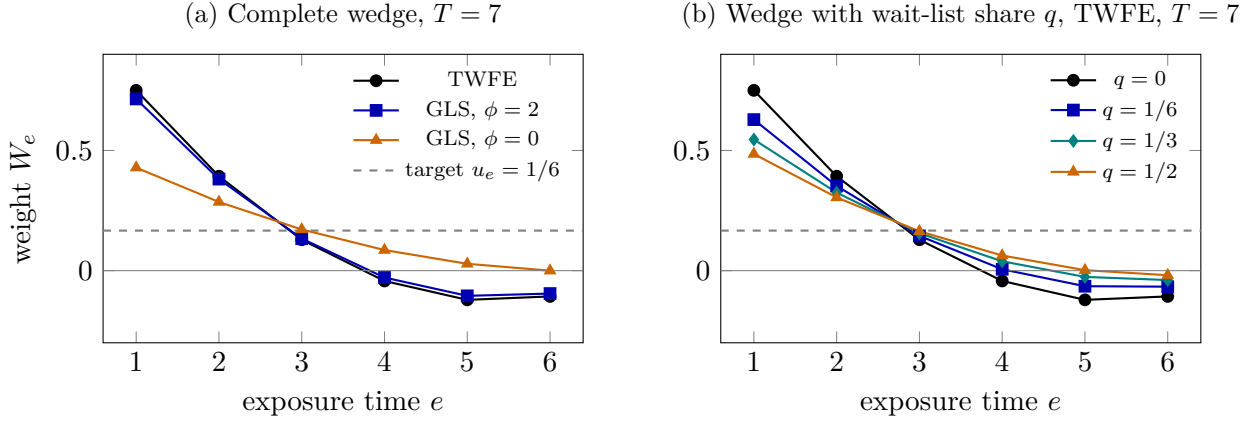
\begin{figure}[t]
\centering
\begin{tikzpicture}
\begin{axis}[
  width=0.48\textwidth, height=5.4cm,
  xlabel={exposure time $e$}, ylabel={weight $W_e$},
  title={\small (a) Complete wedge, $T=7$},
  xmin=0.6, xmax=6.4, ymin=-0.3, ymax=0.9, xtick={1,...,6},
  legend style={font=\scriptsize, at={(0.98,0.98)}, anchor=north east, draw=none, fill=none},
  every axis plot/.append style={thick}]
\addplot[black, mark=*, samples at={1,...,6}] {6*(7-x)*(48-13*x)/1680};
\addlegendentry{TWFE}
\addplot[blue!70!black, mark=square*, samples at={1,...,6}] {(8/8.4)*6*(7-x)*(48-13*x)/1680 + (1-8/8.4)*6*(7-x)*(x-1)/210};
\addlegendentry{GLS, $\phi=2$}
\addplot[orange!80!black, mark=triangle*, samples at={1,...,6}] {3*(7-x)*(6-x)/210};
\addlegendentry{GLS, $\phi=0$}
\addplot[gray, dashed, domain=0.6:6.4] {1/6};
\addlegendentry{target $u_e=1/6$}
\addplot[gray, thin, domain=0.6:6.4] {0};
\end{axis}
\end{tikzpicture}\hfill
\begin{tikzpicture}
\begin{axis}[
  width=0.48\textwidth, height=5.4cm,
  xlabel={exposure time $e$},
  title={\small (b) Wedge with wait-list share $q$, TWFE, $T=7$},
  xmin=0.6, xmax=6.4, ymin=-0.3, ymax=0.9, xtick={1,...,6},
  legend style={font=\scriptsize, at={(0.98,0.98)}, anchor=north east, draw=none, fill=none},
  every axis plot/.append style={thick}]
\addplot[black, mark=*, samples at={1,...,6}] {6*(7-x)*(48-13*x)/(336*5)};
\addlegendentry{$q=0$}
\addplot[blue!70!black, mark=square*, samples at={1,...,6}] {6*(7-x)*(48-13*x+7*x/6)/(336*(5+7/6))};
\addlegendentry{$q=1/6$}
\addplot[teal, mark=diamond*, samples at={1,...,6}] {6*(7-x)*(48-13*x+7*x/3)/(336*(5+7/3))};
\addlegendentry{$q=1/3$}
\addplot[orange!80!black, mark=triangle*, samples at={1,...,6}] {6*(7-x)*(48-13*x+7*x/2)/(336*(5+7/2))};
\addlegendentry{$q=1/2$}
\addplot[gray, dashed, domain=0.6:6.4] {1/6};
\addplot[gray, thin, domain=0.6:6.4] {0};
\end{axis}
\end{tikzpicture}
\caption{Exposure weights of the IT estimator with $T=7$ periods, the geometry of the trial in \cref{sec:example}. (a) Balanced complete wedge under TWFE (\cref{thm:twfe}) and Hussey--Hughes GLS with $\phi=2$ and $\phi=0$ (\cref{thm:gls}). (b) Balanced wedge with wait-list share $q$ under TWFE (\cref{thm:hybrid}). The dashed line is the uniform weight of the target. Curves are drawn directly from the closed forms.}
\label{fig:weights}
\end{figure}

The weights~\eqref{eq:Wtwfe} are the $\psi\to1$ limit of Theorem~1 of \citet{kenny2022analysis}; the new content is the closed form of $\Gamma$ and the resulting bound. The radius of a balanced complete wedge analysed by TWFE is therefore essentially a constant, about $0.12$, whatever the number of periods, steps or clusters (\cref{fig:weights,fig:radius}).

\subsection{The balanced wedge under the Hussey--Hughes model}

The GLS estimator combines the within-cluster (TWFE) estimator with a between-cluster estimator. The following decomposition is classical for balanced panels \citep{maddala1971use}; we state the form we need, which allows wait-listed sequences.

\begin{lemma}[Within--between decomposition]\label{lem:wb}
Consider any design in which every cluster is observed in every period, with covariance $V=I+\phi J$. (i) For every $\phi\ge0$ the GLS estimates of the period effects equal the period means of $\bar Y_{ct}-D_{g(c)t}\hat\beta$, so the estimator acts on period-centred data. (ii) Write $\tilde D_{ct}$ for the two-way residual of $D$, $S_W=\sum_{c,t}\tilde D_{ct}^2$, and $S_B=T\sum_c(\bar D_c-\bar D)^2$, where $\bar D_c$ is the time average of $D$ in cluster $c$ and $\bar D$ the grand mean. Then
\[
\betahat_{\mathrm{GLS}}(\phi)=\frac{S_W\,\betahat_W+\lambda S_B\,\betahat_B}{S_W+\lambda S_B},\qquad \lambda=\frac1{1+T\phi},
\]
where $\betahat_W$ is TWFE-IT and $\betahat_B$ is the regression of period-centred cluster means on $\bar D_c-\bar D$. Hence $W(\phi)=\omega\Wtw+(1-\omega)\Wbt$ with $\omega=S_W/(S_W+\lambda S_B)$, which increases to $1$ as $\phi\to\infty$. (iii) For a vector of treatment regressors, as in ETI, the same holds with $S_W$ and $S_B$ replaced by information matrices.
\end{lemma}

\begin{theorem}[Hussey--Hughes GLS, balanced complete wedge]\label{thm:gls}
For $\mathcal{W}_T$ with $T\ge3$ and $\phi\ge0$:
\begin{enumerate}[label=(\roman*)]
\item the between weights are $\Wbt_e=6(T-e)(e-1)/\{T(T-1)(T-2)\}$, which are nonnegative and zero at $e=1$, and
\[
\omega(\phi)=\frac{T+1}{(T+1)+(T-1)\lambda};
\]
the resulting weights $W(\phi)=\omega\Wtw+(1-\omega)\Wbt$ coincide with equation~(5) of \citet{kenny2022analysis} with $\psi=\phi/(1+\phi)$;
\item
\begin{equation}\label{eq:G2gls}
\Gamma^2(T,\phi)=\frac{\splitfrac{(T+1)(19T^3-31T^2-2T+12)\phi^2}{{}+2(17T^3-20T^2-19T+18)\phi+4(2T-1)(2T-3)}}{5T(T-2)\{(T+1)\phi+2\}^2};
\end{equation}
\item $\Gamma(T,\phi)$ is strictly increasing in $\phi$, so the radius strictly decreases as the intracluster correlation or the cluster-period size increases;
\item $\Gamma^2(T,0)=(2T-1)(2T-3)/\{5T(T-2)\}>4/5$, and therefore $0.1194<\kstar<0.2604$ for every $T\ge4$ and $\phi\ge0$.
\end{enumerate}
\end{theorem}

At $\phi=0$ the weights reduce to $W_e(0)=3(T-e)(T-1-e)/\{T(T-1)(T-2)\}$, which are nonnegative and zero at $e=L$, in line with the observation of \citet{kenny2022analysis} that working independence avoids negative weights. The between-cluster comparisons are made within periods, between treated and untreated clusters; they are concurrent comparisons, and they are what raises the radius at low $\phi$. As $\phi$ grows the estimator relies on within-cluster comparisons across periods and the radius falls to the TWFE value. For $\phi\ge0.5$ and $4\le T\le10$ the radius lies between $0.12$ and $0.16$ (\cref{fig:radius}a; \cref{tab:kgrid}).

\subsection{Linear learning curves and smooth heterogeneity}

The worst case in \cref{prop:worstbias} is a profile proportional to $W-u$, which may look artificial. For complete wedges the worst case over smooth profiles is the ordinary learning curve.

\begin{proposition}[Linear learning curves]\label{prop:ramp}
Consider $\mathcal{W}_T$ under GLS with $\phi\ge0$, or under TWFE ($\omega=1$), and let $\Theta_M=\{\theta:|\theta_{e+1}-\theta_e|\le M,\ e=1,\dots,L-1\}$.
\begin{enumerate}[label=(\roman*)]
\item The \emph{effective exposure} $\ebar=\sum_ee\,W_e$ equals
\[
\ebar(\phi)=\frac{\omega}{2}+(1-\omega)\frac{T+1}{2}\in\Big[\frac12,\frac{T+1}{4}\Big],
\]
with $\ebar=1/2$ under TWFE for every $T$. If $\theta_e=\theta_1+b(e-1)$, then $\E\betahat=\theta_1+b(\ebar-1)$, the linear curve evaluated at $\ebar$, and the bias for $\tau$ is $-\tfrac{b}{2}(\omega T-1)$; under TWFE it is $-bL/2$.
\item The partial sums $\sum_{e\le k}(W_e-1/L)$ are positive for $k=1,\dots,L-1$. Consequently
\[
\sup_{\theta\in\Theta_M}\big|\E\betahat-\tau(\theta)\big|=\frac M2(\omega T-1),
\]
attained only by linear ramps with slope $\pm M$.
\item Under TWFE the linear ramp attains the fraction $\sqrt3\,L/\{(L^2-1)^{1/2}\Gamma\}$ of the worst case over $\Theta(s)$ with the same RMS spread; this fraction is $1$ at $T=3$, $0.95$ at $T=5$, $0.93$ at $T=7$ and tends to $(15/19)^{1/2}=0.89$.
\end{enumerate}
\end{proposition}

Part (i) gives a simple reading of the IT estimand. Under TWFE, the IT estimator of a balanced complete wedge estimates where a linear learning curve stood half a period after implementation, whatever the length of the trial; TATE is the curve at exposure $T/2$. If the effect on the outcome scale rises by $b$ per period, TWFE-IT understates TATE by $bL/2$, which is $L/\{2(L-1)\}$ times the total rise of the effect over the trial. With $T=5$ and a ramp from $0.5\delta$ to $1.5\delta$, the TWFE bias is $-0.67\delta$; the Hussey--Hughes value at $\phi=2$ is $-0.62\delta$ (\cref{tab:curves}).

\subsection{Other levers within complete designs}

The radius of a complete design can be changed by allocating clusters unequally across start periods, or by adding periods. Under TWFE the first lever has exactly no effect when $T=3$.

\begin{proposition}[Allocation invariance at $T=3$]\label{prop:t3}
For a complete design with $T=3$ and any shares $p_2=p$, $p_3=1-p$ with $0<p<1$, the TWFE-IT weights are $W=(3/2,-1/2)$, so $\Gamma=2$ and $\kstar=0.1164$.
\end{proposition}

\begin{numresult}[Reallocation within complete designs]\label{nr:realloc}
With $G=24$ and $T=5$, the highest radius among complete allocations whose IT variance does not exceed that of the balanced wedge is $0.145$ at $\phi=2$ and $0.132$ at $\phi=10$, against $0.133$ and $0.126$ for the balanced wedge. The highest radius at any variance is $0.158$ and $0.144$, at $3.9$ and $4.0$ times the variance. Exhaustive search over integer allocations with $G=60$ ($T=4,5$) and $G=36$ ($T=6$) gives, at equal or lower variance, $0.143$--$0.150$ at $\phi=2$ and $0.126$--$0.135$ under TWFE; the maximisers at any variance place nearly all clusters in the last step and reach $0.146$--$0.163$ at $\phi=2$ at $3$--$9$ times the variance.
\end{numresult}

\begin{conjecture}[Reallocation ceiling]\label{conj:ceiling}
For complete designs with $\phi\ge2$, no allocation with IT variance at most that of the balanced wedge has $\kstar$ above $0.16$, and the supremum of $\kstar$ over all complete designs is approached as the share of the last step tends to one.
\end{conjecture}

The maximisers in \cref{nr:realloc} imitate a design in which most clusters stay untreated until the final period, which is close to a wait-list. Adding periods behaves as \cref{cor:final} predicts. One or two extra baseline periods raise the radius to $0.149$ and $0.161$ at $\phi=2$ with 8--13\% lower IT variance. One or two extra periods after all clusters are treated lower it to $0.117$ and $0.111$, partly because the target then extends to longer exposures, and raise the ETI cost to $3.1$ and $4.0$ times the IT variance of the same design (\cref{tab:levers}).

\section{Concurrent controls}\label{sec:hybrid}

\subsection{Parallel layout}

\begin{proposition}[Parallel design with baseline]\label{prop:parallel}
For $\mathcal{P}_T(p)$ with $0<p<1$ and every $\phi\ge0$, including TWFE, the IT exposure weights are uniform, $W_e=1/(T-1)$. Hence $\Gamma=0$, the IT estimator is unbiased for $\tau$ under every exposure-time profile, and $r^\dagger=\infty$.
\end{proposition}

\citet{lee2024cluster} prove the equivalent statement that in parallel designs with and without baseline the IT estimator coincides with the time-averaged exposure-time estimator. We include a short proof based on \cref{lem:wb} because the same calculation gives \cref{thm:hybrid}.

\subsection{Stepped wedge with a wait-list}

\begin{theorem}[TWFE, balanced wedge with wait-list]\label{thm:hybrid}
For $\mathcal{H}_T(q)$ with $T\ge3$ and $0\le q<1$, let $\pi_q=Tq/(T-2+Tq)$ and $V_e=6e(T-e)/\{T(T-1)(T+1)\}$.
\begin{enumerate}[label=(\roman*)]
\item The TWFE-IT exposure weights are
\[
W_e(q)=\frac{6(T-e)\{(T^2-1)-(2T-1)e+Tqe\}}{T(T-1)(T+1)(T-2+Tq)}=(1-\pi_q)\,\Wtw_e+\pi_q\,V_e .
\]
The kernel $V$ is a probability vector, symmetric about $e=T/2$, the mean exposure of the target.
\item
\[
\Gamma^2(T,q)=\frac{(T-2)\{T^2(T-3)q^2-2T(2T-1)(T-3)q+19T^3-31T^2-2T+12\}}{5T(T+1)(T-2+Tq)^2},
\]
and
\[
\frac{\partial\Gamma^2}{\partial q}=\frac{6(T-2)(T-1)\{q(T^2-3T)-(7T^2-7T-2)\}}{5(T+1)(T-2+Tq)^3}<0,
\]
so the radius increases strictly with the wait-list share.
\item No weight is negative if and only if $q\ge(T-2)/T$.
\item The effective exposure is $\ebar(q)=\{(T-2)+T^2q\}/\{2(T-2+Tq)\}$. Under a linear profile with slope $b$ the bias is $-(1-\pi_q)\,bL/2$: a wait-list share $q$ multiplies the linear-ramp bias of the complete wedge by $(T-2)/(T-2+Tq)$.
\item As $q\to1$, $\Gamma^2\to(T-2)(4T+3)/\{5T(T+1)\}$, so within this family the TWFE-IT radius stays below $0.378$ for $T\ge4$, and its supremum over $q$ tends to $0.2603$ as $T\to\infty$.
\end{enumerate}
\end{theorem}

\begin{figure}[t]
\centering
\begin{tikzpicture}[declare function={
  G2(\T,\p) = ((\T+1)*(19*\T^3-31*\T^2-2*\T+12)*\p^2 + 2*(17*\T^3-20*\T^2-19*\T+18)*\p + 4*(2*\T-1)*(2*\T-3)) / (5*\T*(\T-2)*((\T+1)*\p+2)^2);
  H2(\T,\q) = (\T-2)*(\T^2*(\T-3)*\q^2 - 2*\T*(2*\T-1)*(\T-3)*\q + 19*\T^3-31*\T^2-2*\T+12) / (5*\T*(\T+1)*(\T-2+\T*\q)^2);
}]
\begin{axis}[
  width=0.48\textwidth, height=5.6cm, xmode=log,
  xlabel={$\phi=m\rho/(1-\rho)$}, ylabel={radius $\kappa^*$},
  title={\small (a) Complete wedge, Hussey--Hughes GLS},
  xmin=0.01, xmax=100, ymin=0.10, ymax=0.27,
  legend style={font=\scriptsize, at={(0.98,0.98)}, anchor=north east, draw=none, fill=none},
  every axis plot/.append style={thick, samples=80, domain=0.01:100}]
\addplot[black] {0.23285/sqrt(G2(4,x))}; \addlegendentry{$T=4$}
\addplot[blue!70!black, dashed] {0.23285/sqrt(G2(5,x))}; \addlegendentry{$T=5$}
\addplot[teal, dotted] {0.23285/sqrt(G2(7,x))}; \addlegendentry{$T=7$}
\addplot[orange!80!black, dashdotted] {0.23285/sqrt(G2(10,x))}; \addlegendentry{$T=10$}
\addplot[gray, thin, densely dashed, domain=0.01:100, samples=2] {0.1194};
\end{axis}
\end{tikzpicture}\hfill
\begin{tikzpicture}[declare function={
  H2(\T,\q) = (\T-2)*(\T^2*(\T-3)*\q^2 - 2*\T*(2*\T-1)*(\T-3)*\q + 19*\T^3-31*\T^2-2*\T+12) / (5*\T*(\T+1)*(\T-2+\T*\q)^2);
}]
\begin{axis}[
  width=0.48\textwidth, height=5.6cm,
  xlabel={wait-list share $q$},
  title={\small (b) Wedge with wait-list, TWFE},
  xmin=0, xmax=0.9, ymin=0.10, ymax=0.40,
  legend style={font=\scriptsize, at={(0.02,0.98)}, anchor=north west, draw=none, fill=none},
  every axis plot/.append style={thick, samples=60, domain=0:0.9}]
\addplot[black] {0.23285/sqrt(H2(4,x))}; \addlegendentry{$T=4$}
\addplot[blue!70!black, dashed] {0.23285/sqrt(H2(5,x))}; \addlegendentry{$T=5$}
\addplot[teal, dotted] {0.23285/sqrt(H2(7,x))}; \addlegendentry{$T=7$}
\addplot[orange!80!black, dashdotted] {0.23285/sqrt(H2(10,x))}; \addlegendentry{$T=10$}
\end{axis}
\end{tikzpicture}
\caption{Adequacy radius $\kappa^*$ at 80\% planned power. (a) Balanced complete wedge under Hussey--Hughes GLS, as a function of $\phi$ (\cref{thm:gls}); the dashed line is the TWFE limit $0.1194$ as $T\to\infty$. (b) Balanced wedge with wait-list share $q$ under TWFE (\cref{thm:hybrid}). Curves are drawn from the closed forms.}
\label{fig:radius}
\end{figure}
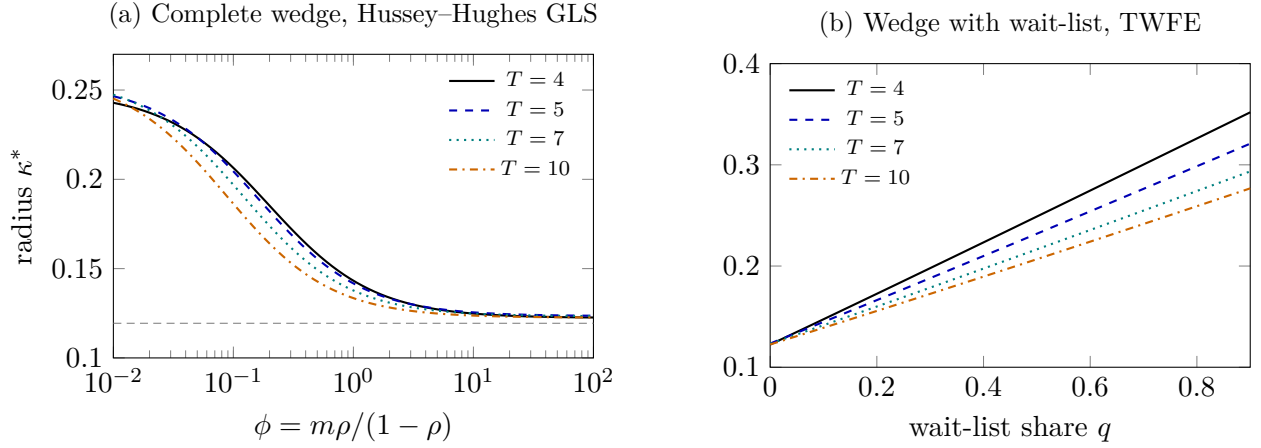

The mixture form in (i) shows what the wait-list does. The IT estimator now combines the complete-wedge contrasts, which compare exposures within and across periods, with contrasts of treated clusters against concurrent wait-listed controls, whose exposure weights $V_e$ are centred on the target. Part (v) is specific to this family: the staggered part still contrasts exposures against each other, so the IT radius stays bounded however large $q$ is. Other hybrid allocations do better (\cref{nr:etiopt}). \Cref{tab:levers} and \cref{fig:radius}b show the size of the effect. Under TWFE with $T=5$, $\kstar$ rises from $0.123$ to $0.159$, $0.177$, $0.195$ and $0.232$ at $q=1/6,1/4,1/3,1/2$; with $T=8$ the values are $0.152$, $0.167$, $0.182$ and $0.212$.

\subsection{Hybrids under the Hussey--Hughes model}

\begin{theorem}[Hussey--Hughes GLS, balanced wedge with wait-list]\label{thm:hybridgls}
For $\mathcal{H}_T(q)$ with $T\ge3$, $0\le q<1$ and $0\le\phi\le\infty$ (with $\phi=\infty$ meaning TWFE), let
\[
t(q,\phi)=\frac{(1-q)+(2T-1-qT)\,\phi}{(T-1)\{(1+q)+(T+1)\phi\}},
\qquad t(q,\infty)=\frac{2T-1-qT}{(T-1)(T+1)} .
\]
\begin{enumerate}[label=(\roman*)]
\item The IT-GLS exposure weights lie on a single one-parameter curve:
\[
W_e=\frac{6\,(T-e)(1-t\,e)}{T(T-1)\{3-(T+1)t\}},\qquad e=1,\dots,T-1,\quad t=t(q,\phi)\in\Big[0,\tfrac{3}{T+1}\Big).
\]
\item Against the TATE target,
\[
\Gamma^2=\frac{(T-2)\{(T-3)(T+1)\,t^2+15\}}{5T\{3-(T+1)t\}^2},
\]
which is strictly increasing in $t$ on $[0,3/(T+1))$ for every $T\ge3$.
\item $t(q,\phi)$ is strictly decreasing in $q$ and strictly increasing in $\phi$. Hence the radius $\kstar$ of the IT-GLS analysis is strictly increasing in the wait-list share and strictly decreasing in the intracluster correlation (equivalently in $\phi$), for every $T\ge3$.
\item Over the whole family, $\Gamma^2>(T-2)/(3T)$, so $\kstar<0.2329\{3T/(T-2)\}^{1/2}$, which is $0.521$ at $T=5$, $0.477$ at $T=7$ and tends to $0.403$. The infimum is approached as $q\to1$ with $\phi=0$, where $W_e\propto T-e$, the design-average exposure distribution.
\item The effective exposure is $\ebar=\sum_e eW_e=(T+1)(2-Tt)/[2\{3-(T+1)t\}]$, so under a linear profile with slope $b$ the bias for $\tau$ is $-b(T-2)/[2\{3-(T+1)t\}]$, whose size is also strictly increasing in $t$.
\end{enumerate}
\end{theorem}

\Cref{thm:hybridgls} shows that the wait-list share and the correlation act on the IT estimand through one scalar. Every design in the family, under every working correlation, puts weights $(T-e)(1-te)$ on exposure $e$ up to normalisation; adding wait-listed clusters or lowering the correlation lowers $t$ and moves the weights towards the design-average shape $T-e$, which is closer to the uniform target. Within this family the IT radius is therefore bounded however large the wait-list (part (iv)); designs outside it, such as the front-loaded hybrids of \cref{nr:etiopt}, do better. Appendix~\ref{app:balanced-waitlist-monotonicity} proves the corresponding monotonicity for the distance to the design-average target, which in this notation is the statement that $t$ is monotone; part (ii) adds that the distance to TATE is monotone in $t$.

With $G=24$ and $T=5$, wait-listing $4$, $8$, $12$ and $16$ clusters (the rest balanced over the four steps) gives $\kstar=0.174$, $0.214$, $0.254$, $0.293$ at $\phi=2$ and $0.162$, $0.199$, $0.237$, $0.274$ at $\phi=10$, each above the corresponding TWFE value, as \cref{thm:hybridgls}(iii) requires. The IT variance is $0.92$--$0.96$ times that of the complete wedge with up to a third wait-listed, and exceeds it from a half wait-listed onwards ($1.04$--$1.08$; \cref{tab:levers}).

Wait-listing one cluster in six thus raises the radius more than any reallocation in \cref{nr:realloc}, at lower IT variance. By \cref{prop:identification}(b) the wait-list has a second effect that no reallocation can have: it makes the final-period effects identifiable without assuming that the effect depends on exposure time only.

\begin{table}[t]
\centering
\begin{threeparttable}
\caption{Design levers with $G=24$ clusters and $T=5$ periods. Radius $\kappa^*$ of the IT analysis at 80\% planned power, and variance relative to the IT analysis of the balanced complete wedge, under Hussey--Hughes GLS.}
\label{tab:levers}
\small
\begin{tabular}{@{}lcccc@{}}
\toprule
 & \multicolumn{2}{c}{$\phi=2$} & \multicolumn{2}{c}{$\phi=10$}\\
\cmidrule(lr){2-3}\cmidrule(lr){4-5}
Design and analysis & $\kappa^*$ & variance & $\kappa^*$ & variance\\
\midrule
\multicolumn{5}{@{}l}{\emph{Complete designs, IT analysis}}\\
Balanced wedge (6 clusters per step) & 0.133 & 1.00 & 0.126 & 1.00\\
More clusters per step & unchanged & lower & unchanged & lower\\
Best reallocation, variance $\le$ balanced & 0.145 & $\le1.00$ & 0.132 & $\le1.00$\\
Best reallocation, any variance & 0.158 & 3.89 & 0.144 & 4.00\\
One extra baseline period & 0.149 & 0.92 & 0.143 & 0.90\\
Two extra baseline periods & 0.161 & 0.87 & 0.157 & 0.85\\
One extra all-treated period\tnote{a} & 0.117 & 0.92 & --- & ---\\
Two extra all-treated periods\tnote{a} & 0.111 & 0.87 & --- & ---\\
\addlinespace
\multicolumn{5}{@{}l}{\emph{Concurrent controls, IT analysis}}\\
4 of 24 wait-listed (5 per step) & 0.174 & 0.92 & 0.162 & 0.93\\
8 of 24 wait-listed (4 per step) & 0.214 & 0.93 & 0.199 & 0.96\\
12 of 24 wait-listed (3 per step) & 0.254 & 1.04 & 0.237 & 1.08\\
16 of 24 wait-listed (2 per step) & 0.293 & 1.34 & 0.274 & 1.40\\
ETI-optimal hybrid\tnote{b} & 0.420 & 1.09 & 0.321 & 1.13\\
IT-optimal design (constant effect)\tnote{c} & 0.187 & 0.91 & 0.173 & 0.93\\
\addlinespace
\multicolumn{5}{@{}l}{\emph{Robust analysis (unbiased for TATE, $\kappa^*=\infty$)}}\\
ETI on balanced wedge & $\infty$ & 2.27 & $\infty$ & 2.51\\
ETI on IT-optimal design\tnote{c} & $\infty$ & 1.47 & $\infty$ & 1.64\\
ETI on ETI-optimal hybrid\tnote{b} & $\infty$ & 1.21 & $\infty$ & 1.34\\
Parallel with baseline, 12/12 (IT $=$ ETI) & $\infty$ & 1.46 & $\infty$ & 1.76\\
\bottomrule
\end{tabular}
\begin{tablenotes}\footnotesize
\item[a] Base design with 4 clusters per step; the TATE horizon lengthens with the added periods. ETI variance relative to IT in the same design: 3.06 and 3.98.
\item[b] Clusters starting in periods 2/3/4/5 and wait-listed: 8/4/1/0 and 11 at $\phi=2$; 7/4/2/0 and 11 at $\phi=10$ (no rollout cap).
\item[c] Allocation minimising the IT variance under a constant effect in the same design space: 6/4/4/5 and 5 at $\phi=2$; 5/5/4/5 and 5 at $\phi=10$.
\end{tablenotes}
\end{threeparttable}
\end{table}

\section{The cost of robust analysis}\label{sec:cost}

\subsection{ETI cost and the robustness--precision frontier}

The radius measures how much the IT analysis tolerates. The alternative is to remove the bias by analysis. For a candidate design $d$ we measure the cost of doing so by
\[
\rho^2_d=\frac{\Var(\tauhat;d)}{\Var(\betahat_{\mathrm{IT}};\mathcal{W}_T)},
\]
the variance of the ETI estimator in $d$ relative to the IT estimator in the balanced complete wedge with the same number of clusters. Between the two analyses lies a continuum of linear estimators that trade bias for variance.

\begin{definition}[Robustness--precision frontier]\label{def:frontier}
For a design $d$ with known covariance, let $F_d(\gamma)$ be the minimum variance of a period-invariant linear estimator with $\one^\top W=1$ and $\sqrt L\lVert W-u\rVert_2\le\gamma$.
\end{definition}

\begin{proposition}[Frontier]\label{prop:frontier}
$F_d$ is convex and non-increasing. If TATE is identified, $F_d(0)=\Var(\tauhat;d)$ for the GLS fit of the ETI model, and $F_d(\gamma)=\Var(\betahat_{\mathrm{IT\text{-}GLS}};d)$ for $\gamma\ge\Gamma(\betahat_{\mathrm{IT\text{-}GLS}})$.
\end{proposition}

The frontier summarises a design for a planner who does not yet know which analysis to commit to: its two ends are the IT and ETI analyses, and its interior is traced by penalised estimators $\min\{\Var+\nu\cdot\text{bias}^2\}$, which are the bias-aware estimators of \citet{armstrong2018optimal} for the class $\Theta(s)$. Concurrent controls lower the left end of the frontier much more than the right end, as the following results show.

\subsection{Wait-listing under a rollout cap}

Stepped wedges are often chosen because only a limited number of clusters can start the intervention in each period. We therefore compare designs under the same rollout capacity as the balanced wedge.

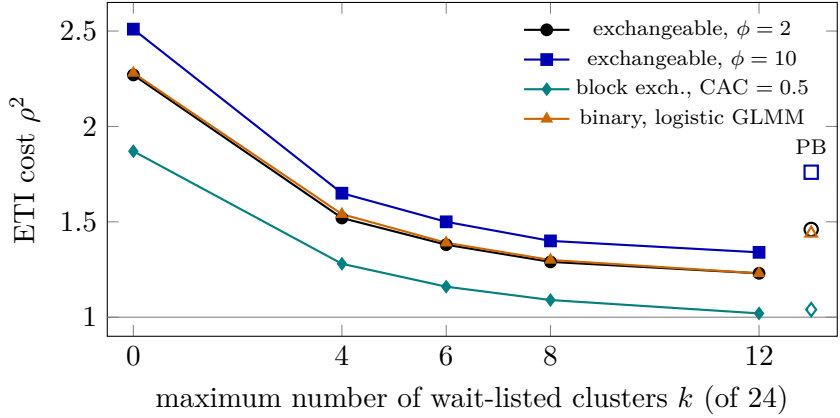
\begin{figure}[t]
\centering
\begin{tikzpicture}
\begin{axis}[
  width=0.68\textwidth, height=6cm,
  xlabel={maximum number of wait-listed clusters $k$ (of 24)},
  ylabel={ETI cost $\rho^2$},
  xmin=-0.5, xmax=13.5, ymin=0.9, ymax=2.65, xtick={0,4,6,8,12},
  legend style={font=\scriptsize, at={(0.98,0.98)}, anchor=north east, draw=none, fill=none},
  every axis plot/.append style={thick}]
\addplot[black, mark=*] coordinates {(0,2.27) (4,1.52) (6,1.38) (8,1.29) (12,1.23)};
\addlegendentry{exchangeable, $\phi=2$}
\addplot[blue!70!black, mark=square*] coordinates {(0,2.51) (4,1.65) (6,1.50) (8,1.40) (12,1.34)};
\addlegendentry{exchangeable, $\phi=10$}
\addplot[teal, mark=diamond*] coordinates {(0,1.87) (4,1.28) (6,1.16) (8,1.09) (12,1.02)};
\addlegendentry{block exch., CAC $=0.5$}
\addplot[orange!80!black, mark=triangle*] coordinates {(0,2.28) (4,1.54) (6,1.39) (8,1.30) (12,1.23)};
\addlegendentry{binary, logistic GLMM}
\addplot[black, only marks, mark=o, mark size=2.5pt] coordinates {(13,1.46)};
\addplot[blue!70!black, only marks, mark=square, mark size=2.5pt] coordinates {(13,1.76)};
\addplot[teal, only marks, mark=diamond, mark size=2.5pt] coordinates {(13,1.04)};
\addplot[orange!80!black, only marks, mark=triangle, mark size=2.5pt] coordinates {(13,1.44)};
\addplot[gray, thin, domain=-0.5:13.5] {1};
\node[font=\scriptsize, anchor=south] at (axis cs:13,1.80) {PB};
\end{axis}
\end{tikzpicture}
\caption{ETI cost $\rho^2$: variance of the ETI estimator of TATE in the best design with at most $k$ wait-listed clusters and at most 6 clusters starting per period, relative to the IT variance of the balanced complete wedge; $G=24$, $T=5$. Hollow markers at the right (PB) are the parallel design with baseline, 12 clusters treated from period 2 and 12 wait-listed. Values from \texttt{frontier\_waitlist.py}.}
\label{fig:etcost}
\end{figure}

\begin{numresult}[ETI cost against the number wait-listed]\label{nr:frontier}
With $G=24$, $T=5$ and at most 6 clusters starting per period, the smallest ETI cost among designs with at most $k$ wait-listed clusters is, under exchangeable correlation with $\phi=2$, $2.27$ ($k=0$), $1.52$ ($k=4$), $1.38$ ($k=6$), $1.29$ ($k=8$) and $1.23$ ($k=12$). The parallel design with baseline (12/12) costs $1.46$. The same pattern holds with $\phi=10$, with block-exchangeable correlation (cluster autocorrelation $0.5$) and for a binary outcome analysed by a linearised logistic mixed model (\cref{fig:etcost}; \cref{tab:frontier} in \cref{app:tables}).
\end{numresult}

Wait-listing 4 of 24 clusters achieves 72\% of the reduction available with up to 12 wait-listed ($2.27\to1.52$, against $1.23$). The best designs with $k\le4$, $6$ and $8$ are $(6,6,2,6)+4$, $(6,6,0,6)+6$ and $(6,6,1,3)+8$ (clusters starting in periods 2/3/4/5, plus those wait-listed): the wait-listed clusters come from the third and fourth steps, the first two steps stay full, and the rollout is kept. With exchangeable correlation a hybrid with a third of clusters wait-listed is cheaper than the parallel design; with weak cluster autocorrelation ($0.5$) the parallel design ($1.04$) is as good as the best hybrid ($1.02$), and the argument for a hybrid then rests on keeping a staggered rollout rather than on precision.

\begin{numresult}[ETI-optimal hybrids]\label{nr:etiopt}
Under the same rollout cap, the design that minimises ETI variance wait-lists 11--13 of 24 clusters (46--54\%) and starts the remainder early: at $\phi=2$ it is $(6,6,1,0)+11$, with ETI cost $1.23$. Across exchangeable ($\phi=2,10$), block-exchangeable (CAC $0.8$, $0.5$), exponential-decay ($r=0.8$, $0.5$), binary-outcome and misspecified-working-model scenarios the ETI cost of these designs is $1.02$--$1.41$, against $1.85$--$2.51$ for ETI in the complete wedge, and their IT radius is $0.31$--$0.55$. Without the cap the optimum wait-lists 10--11 clusters (42--46\%) and the ETI cost changes by at most $0.08$.
\end{numresult}

The ETI-optimal designs are front-loaded hybrids: a short stagger in the first two steps and a large concurrent-control arm. They are close to a parallel design with baseline, but cheaper (at $\phi=2$, $1.21$ against $1.46$ without the cap), which extends the finding of \citet{lee2024cluster} that the parallel design with baseline is more efficient than the stepped wedge for time-averaged effects. They also differ from the precision-optimal design under a constant effect. In the same design space, the allocation that minimises the IT variance, which is the discrete analogue of the optimum of \citet{girling2016statistical}, is $(6,4,4,5)+5$ at $\phi=2$ and $(5,5,4,5)+5$ at $\phi=10$, with or without the cap. It already wait-lists 5 of 24 clusters: its IT variance is $0.91$--$0.93$ times that of the balanced wedge, its IT radius $0.19$ and $0.17$, and its ETI cost $1.47$ and $1.64$. The ETI-optimal design wait-lists about twice as many clusters and starts the rest earlier, which lowers the ETI cost to $1.21$ and $1.34$ but raises the IT variance to $1.09$ and $1.13$ times that of the balanced wedge. Concurrent controls are therefore already favoured when the effect is constant, and exposure-time heterogeneity strengthens the case for them. The large-study result of \citet{girling2016statistical}, a stepped share equal to $R$ ($10/11$ at $T=5$, $\phi=2$), is derived for a continuum of uptake times and allows clusters treated throughout; with five periods, a common baseline period and no always-treated clusters, the discrete optimum places more clusters outside the stagger.

We do not have a proof that ETI cost decreases with the wait-list share in general. By \cref{lem:wb}(iii) the ETI variance under TWFE is $c^\top(X^\top MX)^{-1}c$ with an information matrix that is block-structured in periods and exposures, and \cref{prop:identification}(a) explains why the complete wedge is expensive: $\theta_L$ is reached only through a chain of within-period differences. A closed form for the hybrid family is the natural next step.

\subsection{Inference at a declared level of heterogeneity}

A trial team that can state a plausible size $\kappa$ of exposure-time heterogeneity, as a fraction of the planned effect, can compare analyses by the length of intervals that are valid over $\Theta(\kappa\delta)$. For the IT estimator the bias-aware interval of \citet{armstrong2020simple} has half-length $\mathrm{cv}_\alpha(\bar b/\SD)\,\SD$, where $\bar b=\Gamma\kappa\delta$ and $\mathrm{cv}_\alpha(t)$ is the $1-\alpha$ quantile of $|N(t,1)|$. The ETI interval has half-length $z_{1-\alpha/2}\,\varrho\,\SD$, where $\varrho^2$ is the ETI-to-IT variance ratio in the same design.

\begin{proposition}[Crossover]\label{prop:crossover}
At planned power $80\%$ and $\alpha=0.05$, the bias-aware IT interval is shorter than the ETI interval if and only if $\mathrm{cv}_{0.05}(2.8016\,\Gamma\kappa)<1.96\,\varrho$. Since $\mathrm{cv}_{0.05}$ is increasing, there is a unique crossover $\kappa_\times$.
\end{proposition}

For the balanced wedge with $T=5$ the crossover is $\kappa_\times=0.267$ at $\phi=2$ and $0.281$ at $\phi=10$. Below it a bias-aware IT interval is the shorter honest interval; above it the ETI interval is. The best estimator on the frontier between the two is at most $14\%$ shorter than the better endpoint (at $\kappa=0.3$, $\phi=10$), so in practice the choice is between the two endpoints, and the design determines how expensive the ETI endpoint is.

\begin{table}[t]
\centering
\begin{threeparttable}
\caption{Half-length of the shortest valid interval for TATE at a declared heterogeneity level $\kappa$ (RMS, as a fraction of the planned effect), relative to the bias-aware IT interval in the balanced complete wedge; $G=24$, $T=5$.}
\label{tab:honest}
\small
\begin{tabular}{@{}lcccc@{}}
\toprule
 & \multicolumn{2}{c}{$\phi=2$} & \multicolumn{2}{c}{$\phi=10$}\\
\cmidrule(lr){2-3}\cmidrule(lr){4-5}
Design and analysis & $\kappa=0.2$ & $\kappa=0.3$ & $\kappa=0.2$ & $\kappa=0.3$\\
\midrule
Balanced wedge, bias-aware IT & 1.000 & 1.000 & 1.000 & 1.000\\
Best reallocation of steps, bias-aware IT & 0.968 & 0.956 & 0.979 & 0.968\\
8 of 24 wait-listed, bias-aware IT & 0.841 & 0.802 & 0.847 & 0.808\\
12 of 24 wait-listed, bias-aware IT & 0.847 & 0.788 & 0.853 & 0.793\\
Balanced wedge, ETI & 1.125 & 0.948 & 1.156 & 0.968\\
ETI-optimal hybrid, ETI & 0.823 & 0.694 & 0.844 & 0.708\\
\bottomrule
\end{tabular}
\begin{tablenotes}\footnotesize
\item Bias-aware intervals use the worst-case bias over $\Theta(\kappa\delta)$ \citep{armstrong2020simple}. ETI-optimal hybrid as in \cref{tab:levers}.
\end{tablenotes}
\end{threeparttable}
\end{table}

\Cref{tab:honest} summarises the comparison in one criterion. At a declared $\kappa$ of $0.2$--$0.3$, the best reallocation of steps shortens the valid interval by 2--4\%, wait-listing a third of clusters by 15--20\%, and the ETI-optimal hybrid analysed by ETI by 16--31\%.

\Cref{tab:curves} fixes two effect curves in absolute units rather than taking a worst case. With a linear ramp from $0.5\delta$ to $1.5\delta$ (RMS variation $37\%$ of TATE), the IT analysis of the complete wedge has bias $-0.62\delta$ and covers TATE $59\%$ of the time; with 8 of 24 wait-listed the bias is $-0.34\delta$ and coverage $84\%$. The ETI interval is unbiased in all designs, and with 8 wait-listed it is $25\%$ shorter than in the complete wedge.

\begin{table}[t]
\centering
\begin{threeparttable}
\caption{Bias, coverage of the nominal 95\% interval, standard error and root-mean-square error for two fixed effect curves, in units of the planned effect $\delta$ (80\% power for IT in the balanced wedge); $G=24$, $T=5$, $\phi=2$.}
\label{tab:curves}
\small
\begin{tabular}{@{}llcccccc@{}}
\toprule
 & & \multicolumn{2}{c}{Linear ramp $0.5\delta\to1.5\delta$} & \multicolumn{2}{c}{Ramp to plateau} & & \\
\cmidrule(lr){3-4}\cmidrule(lr){5-6}
Design & Analysis & bias & coverage & bias & coverage & SE & RMSE (ramp)\\
\midrule
Balanced wedge & IT & $-0.62$ & 0.59 & $-0.28$ & 0.88 & 0.36 & 0.72\\
Balanced wedge & ETI & 0 & 0.95 & 0 & 0.95 & 0.54 & 0.54\\
4 of 24 wait-listed & IT & $-0.46$ & 0.73 & $-0.20$ & 0.91 & 0.34 & 0.58\\
4 of 24 wait-listed & ETI & 0 & 0.95 & 0 & 0.95 & 0.44 & 0.44\\
8 of 24 wait-listed & IT & $-0.34$ & 0.84 & $-0.15$ & 0.93 & 0.35 & 0.49\\
8 of 24 wait-listed & ETI & 0 & 0.95 & 0 & 0.95 & 0.41 & 0.41\\
Parallel with baseline & IT & 0 & 0.95 & 0 & 0.95 & 0.43 & 0.43\\
\bottomrule
\end{tabular}
\begin{tablenotes}\footnotesize
\item Wait-list designs under the rollout cap of \cref{nr:frontier}: $(6,6,2,6)+4$ and $(6,6,1,3)+8$. The ramp to plateau is $1-e^{-e}$ rescaled to mean one (RMS variation $16\%$ of TATE). A waning effect gives the same results with the sign of the bias reversed.
\end{tablenotes}
\end{threeparttable}
\end{table}

\subsection{Monte Carlo checks and small numbers of clusters}\label{sec:smallG}

The variances above are computed from the GLS algebra. A brute-force check that fits the full-data GLS model to 4{,}000 simulated trials gives an ETI-to-IT variance ratio of $2.17$ in the balanced wedge (algebra $2.27$) and $1.20$ for the ETI-optimal hybrid (algebra $1.21$) at $\phi=2$. The script draws a new set of simulated trials for each estimator, so each ratio compares variances from two independent sets of 4{,}000 normally distributed estimates. Its Monte Carlo standard error is therefore about 3\%, and both differences are within 1.5 standard errors.

Small-sample corrections do not repair heterogeneity bias. In 20{,}000 simulated balanced wedges with $T=5$, $\phi=2$ and a linear ramp at $\kappa=0.3$, TWFE-IT with the leave-one-cluster-out jackknife variance (CV3) and $t_{G-1}$ critical values covered TATE in $79\%$, $73\%$ and $71\%$ of trials with $G=8$, $16$ and $24$, against $68\%$ predicted for a known variance; the CV3 standard error exceeded the true standard deviation by only $1$--$5\%$, and the gain in coverage at $G=8$ comes from the $t_7$ critical value. ETI with CV3 and $t_{G-1}$ covered $96\%$ (\cref{tab:smallG} in \cref{app:tables}). Variance corrections of this kind, including the CARE workflow \citep{alexeev2026cluster}, address misspecified covariance and few clusters; they do not address bias from exposure-time heterogeneity, which has to be handled by the estimator or the design. Comparisons of small-sample corrections for ETI models are in \citet{ouyang2026which}.

\section{Illustration: withdrawal of weekend allied-health services}\label{sec:example}

\subsection{Trial and data}

\citet{haines2017impact} report two stepped-wedge trials on 12 acute medical and surgical wards in two Melbourne hospitals. In Trial~1 the existing weekend allied-health service was withdrawn. The public patient-level file (supplementary file S2 of the article) contains 14{,}834 Trial~1 admissions in 12 wards and seven study periods. From the second period onwards the service was withdrawn from one ward at each hospital per period, so the trial is a complete stepped wedge with six sequences of two wards, one from each hospital, and $T=7$, $L=6$. There were 8{,}038 admissions before withdrawal and 6{,}796 after. Ward-periods contain 76 to 397 admissions, 177 on average. At the hospital for which admission dates are released, the periods are consecutive months from February to August 2014; dates are not released for the other hospital, so, like \citet{kenny2022analysis}, we index time by study period. The outcome is the natural logarithm of acute length of stay, as in \citet{kenny2022analysis}; a positive effect means longer stays after withdrawal.

\subsection{Reproduction of published analyses}

\Cref{tab:haines_fits} reports our fits. With a random ward intercept (the Hussey--Hughes model), categorical study period and REML, the IT estimate is $0.114$ (95\% CI $0.054$ to $0.174$), the ETI estimate of TATE is $0.229$ ($0.133$ to $0.325$) and the ETI estimate of the long-term effect $\theta_6$ is $0.291$ ($0.107$ to $0.476$). These agree with Figures~7 and~8 of \citet{kenny2022analysis} to the precision with which values can be read from those figures, and so do the six exposure-specific estimates, which rise from $0.144$ at $e=1$ to $0.319$ at $e=5$ and $0.291$ at $e=6$ (\cref{fig:haines}a). \citet[Section 4]{lee2024cluster} report the same re-analysis, with exposure effects $(0.14, 0.15, 0.21, 0.26, 0.32, 0.29)$ and a between-ward variance of $0.08$.

Under the nested-exchangeable and independence structures our Akaike information criteria for the IT and ETI models equal those in Table~1 of \citet{lee2025analysis} to two decimals. For their calendar-time-indicator model, which has treatment-by-period effects for periods 2--6 because the period-7 effect is not identified (\cref{prop:identification}), the published values ($37{,}001.76$ and $37{,}677.40$) are about 6{,}000 points below ours. We reproduce them to two decimals by fitting that model without the 2{,}064 admissions of period 7, that is, to 12{,}770 rather than 14{,}834 observations. The large advantage of the calendar-time model in that table therefore compares criteria computed on different data, and is not evidence about the form of heterogeneity. On the same data, maximum-likelihood AIC, the appropriate basis for comparing fixed-effect structures, is lower for the calendar-time model than for the IT and ETI models by 8--18 points with random ward effects. The calendar-time-averaged estimate ($0.113$ and $0.129$ with random ward effects) is close to the IT estimate, as \citet{lee2025analysis} found. This is a moderate preference from a comparison of two restricted models with 12 clusters. It does not establish which form of heterogeneity is present, and in a complete design the data cannot settle the question without such restrictions (\cref{prop:identification}).

Two further features of \cref{tab:haines_fits} matter for what follows. Fixed and random ward effects give almost the same estimates, because the estimated ward variance is large relative to the within-cell variance. And the leave-one-ward-out jackknife (CV3) standard error of the IT estimate is about twice the model-based one ($0.060$ against $0.030$), whereas for the ETI estimate of TATE the two are closer ($0.070$ against $0.049$). This is consistent with exposure-time variation left in the residuals of the IT model. We did not attempt to reproduce the primary analysis of \citet{haines2017impact}, which used generalised linear mixed models for log length of stay with adjustment for ward-level data from the two preceding years and reported an increase of 1.3 days (95\% CI 0.9 to 1.8); a linear model for untransformed acute length of stay with ward and period fixed effects gives an increase of 1.59 days (CV3 standard error 0.37).

\begin{table}[t]
\centering
\begin{threeparttable}
\caption{Trial 1 of \citet{haines2017impact}, log acute length of stay: IT, ETI and calendar-time-indicator (CTI) analyses with categorical study period. Estimates with model-based 95\% intervals, leave-one-ward-out jackknife (CV3) intervals with $t_{11}$ critical values, and AIC with the published value where one exists.}
\label{tab:haines_fits}
\footnotesize
\setlength{\tabcolsep}{4pt}
\begin{tabular}{@{}lllcccc@{}}
\toprule
Ward effects & Model & Estimand & Estimate (95\% CI) & CV3 95\% CI & AIC & Published AIC\tnote{a}\\
\midrule
Random intercept & IT & IT & 0.114 (0.054, 0.174) & ($-$0.017, 0.245) & 42{,}952.04 & \\
Random intercept & ETI & TATE & 0.229 (0.133, 0.325) & (0.074, 0.384) & 42{,}971.96 & \\
Random intercept & ETI & LTE & 0.291 (0.107, 0.476) & (0.091, 0.492) & 42{,}971.96 & \\
Random intercept & CTI & CTATE & 0.113 (0.051, 0.174) & ($-$0.069, 0.295) & 42{,}948.18 & \\
Nested exchangeable & IT & IT & 0.126 (0.039, 0.213) & ($-$0.004, 0.255) & 42{,}932.77 & 42{,}932.77\\
Nested exchangeable & ETI & TATE & 0.255 (0.116, 0.394) & (0.104, 0.406) & 42{,}953.94 & 42{,}953.94\\
Nested exchangeable & CTI & CTATE & 0.129 (0.047, 0.211) & ($-$0.045, 0.303) & 42{,}938.50 & 37{,}001.76\\
Independence & IT & IT & 0.165 (0.119, 0.210) & ($-$0.271, 0.601) & 43{,}772.71 & 43{,}772.71\\
Independence & ETI & TATE & 0.213 (0.159, 0.267) & ($-$0.485, 0.912) & 43{,}762.17 & 43{,}762.17\\
Independence & CTI & CTATE & 0.155 (0.108, 0.203) & ($-$0.341, 0.651) & 43{,}703.83 & 37{,}677.40\\
Fixed (TWFE) & IT & IT & 0.113 (0.053, 0.173) & ($-$0.014, 0.240) & 42{,}965.95 & \\
Fixed (TWFE) & ETI & TATE & 0.229 (0.132, 0.326) & (0.072, 0.386) & 42{,}986.00 & \\
\bottomrule
\end{tabular}
\begin{tablenotes}\footnotesize
\item[a] \citet[Table 1]{lee2025analysis}. AIC is REML-based for the random-effect models and maximum-likelihood-based for the independence models, the conventions under which the published IT and ETI values are reproduced. The published CTI values are reproduced exactly when period 7 (2{,}064 admissions) is excluded; our CTI values use all admissions.
\item Published estimates: \citet[Figures 7--8]{kenny2022analysis} show, for the random-intercept model, IT about 0.11 (0.05, 0.17), ETI TATE about 0.23 (0.13, 0.33) and ETI LTE about 0.29 (0.11, 0.48), read from the figures. LTE: long-term effect $\theta_6$; CTATE: average of calendar-time effects over periods 2--6.
\end{tablenotes}
\end{threeparttable}
\end{table}

\subsection{Design diagnostic}

The fitted random-intercept model gives $\hat\sigma^2=1.052$ and $\hat\tau^2=0.080$, an intracluster correlation of $0.071$; with 177 admissions per ward-period this is $\hat\phi=13.5$ ($\hat\psi=0.93$). The nested-exchangeable fit gives a cluster autocorrelation of $0.92$. For the trial's actual cell sizes, the IT estimator's exposure weights are $0.744$, $0.395$, $0.123$, $-0.038$, $-0.106$ and $-0.118$ for $e=1,\dots,6$; the weights of the final period sum to zero to machine precision (\cref{lem:balance}). Hence $\Gamma=1.872$ and $\kstar=0.124$. The nested-exchangeable and fixed-effects fits give $0.125$ and $0.123$, and \cref{thm:gls} for the balanced design with equal cells at $\hat\phi$ gives $0.1245$: unequal cell sizes barely change the radius. The effective exposure $\sum_ee\,W_e$ is $0.51$ under the random-intercept fit and $0.49$ under fixed effects, close to the value $1/2$ of \cref{prop:ramp}.

The IT--ETI discrepancy can be read directly in terms of these weights. When the IT and ETI models are fitted with the same covariance matrix, the IT estimate equals $\sum_eW_e\hat\theta_e$ computed from the ETI fit: the IT model is the ETI model with equal exposure effects, and the GLS residuals of the larger model are orthogonal, in the metric of $V^{-1}$, to the columns of the smaller one. Therefore $\hat\beta_{\mathrm{IT}}-\tauhat=(W-u)^\top\hat\theta$ exactly, which estimates the bias of the IT analysis for TATE with a standard error obtained from the ETI covariance. Here it is $-0.115$ (standard error $0.038$): the IT analysis estimates about half of TATE. (With variance components re-estimated in each fit the identity holds to within $10^{-4}$.) The least-squares slope of $\hat\theta_e$ on $e$ is $0.037$ per period (standard error $0.014$), and \cref{prop:ramp} predicts a TWFE bias of $-3\hat b=-0.111$ for a linear curve with that slope, against the observed TWFE difference of $-0.116$.

To put the heterogeneity on the scale of the radius, note that the trial's IT standard error ($0.0305$) implies an effect detectable with 80\% power of $\delta_{80}=2.8016\times0.0305=0.085$ on the log scale, about a 9\% increase in length of stay. The RMS spread of $\hat\theta$ around its mean is $0.067$; subtracting the part expected from estimation error leaves $0.055$, that is $0.64\,\delta_{80}$, about five times the radius ($0.40\,\delta_{80}$ under the nested-exchangeable fit). If $\hat\theta$ were the true curve, the nominal 95\% IT interval would cover TATE in 3\% of trials (17\% under the nested-exchangeable fit). The trial is large, so the effect it was able to detect is small, and the same heterogeneity is large relative to it (\cref{rem:radius}). These statements assume that the effect varies with exposure time. As noted above, the data do not exclude calendar-time variation, under which the IT estimate is close to the calendar-time average.

\subsection{Counterfactual redesign}

We now ask what alternative designs with the same 12 wards and seven periods would have offered. \citet[Section 4]{lee2024cluster} asked a related question with the same trial. They simulated one dataset from the fitted model for each of four fixed designs (stepped wedge, parallel, parallel with baseline, and parallel with baseline plus an all-treated final period), using the trial's cell sizes, and compared the standard errors of exposure-time and constant-effect estimators. Our exercise differs in four ways. It reports the adequacy radius and the IT bias and coverage for each design, not only the precision of the estimators. It searches over all intermediate allocations of wards to start periods and to a wait-list under the trial's rollout capacity, rather than comparing given designs. It computes design variances analytically and checks them with 2{,}000 simulated trials per design. And it is set against the diagnostic for the trial's actual design given above. \emph{The trial's outcomes under these designs were not observed.} All results below are computed from a model calibrated to the trial (analytically, and by simulating trials from it), not from data collected under those designs. The calibration uses the random-intercept variance components estimated above, 177 admissions in every ward-period, and the trial's ETI estimates $\hat\theta$ as the true effect curve, so that TATE equals $0.229$. We keep the trial's rollout capacity of at most two wards withdrawn per period, and enumerate all allocations of wards to start periods 2--7 and to a wait-list (wards that would keep the weekend service until the end of follow-up). D1 and D2 are the allocations with two and four wards wait-listed that minimise the variance of the ETI estimator; D3 minimises it over any number wait-listed. D4 (parallel with baseline) and D5 (ETI-optimal without the cap) do not respect the rollout capacity and are shown for reference. For each design we compute the IT radius, the IT and ETI variances relative to the IT variance of the trial's design, the ETI 95\% half-width, the value the IT analysis would estimate under $\hat\theta$, and the coverage of the nominal IT interval; we also simulate 2{,}000 trials per design and refit both models by REML.

\begin{table}[t]
\centering
\begin{threeparttable}
\caption{Counterfactual redesign of Trial 1 (12 wards, 7 periods): designs compared under a model calibrated to the trial. Outcomes under D1--D5 are not observed; entries are model-based, with Monte Carlo values in parentheses or in the last column.}
\label{tab:haines}
\footnotesize
\setlength{\tabcolsep}{3.5pt}
\begin{tabular}{@{}llccccccc@{}}
\toprule
 & Wards starting in & IT & IT & ETI & ETI 95\% & IT expected & IT & ETI\\
Design & periods 2--7 + wait-list & $\kappa^*$ & variance & variance & half-width & estimate & coverage & coverage\\
\midrule
D0 trial design & 2/2/2/2/2/2 + 0 & 0.124 & 1.00 & 2.67 & 0.095 & 0.113 & 0.03 (0.03) & 0.96\\
D1 2 wait-listed & 2/2/2/0/2/2 + 2 & 0.159 & 1.00 & 1.86 & 0.079 & 0.138 & 0.14 (0.15) & 0.95\\
D2 4 wait-listed & 2/2/2/1/0/1 + 4 & 0.212 & 1.07 & 1.60 & 0.074 & 0.161 & 0.40 (0.39) & 0.96\\
D3 ETI-optimal & 2/2/2/1/0/0 + 5 & 0.257 & 1.14 & 1.55 & 0.073 & 0.171 & 0.56 (0.54) & 0.95\\
\addlinespace
D4 parallel\tnote{a} & 6/0/0/0/0/0 + 6 & $\infty$ & 2.46 & 2.46 & 0.091 & 0.229 & 0.95 (0.94) & 0.94\\
D5 ETI-optimal\tnote{a} & 3/2/1/1/0/0 + 5 & 0.262 & 1.23 & 1.54 & 0.072 & 0.173 & 0.60 (0.58) & 0.95\\
\bottomrule
\end{tabular}
\begin{tablenotes}\footnotesize
\item Calibration: random ward intercept with $\hat\sigma^2=1.052$, $\hat\tau^2=0.080$, 177 admissions per ward-period, effect curve $\hat\theta$ of the trial's ETI fit (TATE $0.229$). Variances relative to IT in D0. ``IT expected estimate'' is $\sum_eW_e\hat\theta_e$; IT coverage is that of the nominal 95\% interval for TATE, analytic (Monte Carlo); ETI coverage is the Monte Carlo coverage of the model-based 95\% interval. Monte Carlo: 2{,}000 simulated trials per design, REML refits.
\item[a] Exceeds the rollout capacity of two wards per period.
\end{tablenotes}
\end{threeparttable}
\end{table}

\begin{figure}[t]
\centering
\begin{tikzpicture}
\begin{axis}[
  width=0.47\textwidth, height=5.6cm,
  xlabel={exposure time $e$ (periods)}, ylabel={effect on log length of stay},
  title={\small (a) Trial 1: ETI estimates and IT estimate},
  xmin=0.5, xmax=6.5, ymin=-0.05, ymax=0.52, xtick={1,...,6},
  legend style={font=\scriptsize, at={(0.02,0.98)}, anchor=north west, draw=none, fill=none}]
\addplot[only marks, mark=*, black, error bars/.cd, y dir=both, y explicit]
  coordinates {(1,0.144) +- (0,0.068) (2,0.147) +- (0,0.081) (3,0.213) +- (0,0.099)
               (4,0.260) +- (0,0.118) (5,0.319) +- (0,0.147) (6,0.291) +- (0,0.185)};
\addlegendentry{ETI $\hat\theta_e$ (95\% CI)}
\addplot[blue!70!black, thick, dashed, domain=0.5:6.5, samples=2] {0.229};
\addlegendentry{ETI TATE $0.229$}
\addplot[orange!80!black, very thick, dashdotted, domain=0.5:6.5, samples=2] {0.114};
\addlegendentry{IT $0.114$}
\addplot[gray, thin, domain=0.5:6.5, samples=2] {0};
\end{axis}
\end{tikzpicture}\hfill
\begin{tikzpicture}
\begin{axis}[
  width=0.47\textwidth, height=5.6cm,
  xlabel={IT radius $\kappa^*$}, ylabel={ETI 95\% half-width},
  title={\small (b) Counterfactual designs (model-based)},
  xmin=0.10, xmax=0.32, ymin=0.068, ymax=0.100,
  xtick={0.10,0.15,0.20,0.25,0.30},
  yticklabel style={/pgf/number format/fixed, /pgf/number format/precision=3},
  every node near coord/.append style={font=\scriptsize}]
\addplot[only marks, mark=*, black, nodes near coords, point meta=explicit symbolic,
  every node near coord/.append style={anchor=south west}]
  coordinates {(0.124,0.095) [D0] (0.159,0.079) [D1] (0.212,0.074) [D2] (0.257,0.073) [D3]};
\addplot[only marks, mark=*, black, nodes near coords, point meta=explicit symbolic,
  every node near coord/.append style={anchor=north west}]
  coordinates {(0.262,0.072) [D5]};
\addplot[only marks, mark=o, black, nodes near coords, point meta=explicit symbolic,
  every node near coord/.append style={anchor=east}]
  coordinates {(0.31,0.091) [D4 ($\kappa^*=\infty$)]};
\end{axis}
\end{tikzpicture}
\caption{Weekend allied-health disinvestment trial (Trial~1 of \citealp{haines2017impact}). (a) Exposure-specific ETI estimates (random ward intercept, REML) with 95\% intervals, the ETI estimate of TATE and the IT estimate; the IT estimate equals $\sum_eW_e\hat\theta_e$ with the weights given in the text. (b) Designs with the same 12 wards and 7 periods, placed by the radius of their IT analysis and the half-width of their ETI interval under the model calibrated to the trial; D0 is the trial's design. D4 has $\kappa^*=\infty$ and is drawn at the right edge. Outcomes under D1--D5 are not observed.}
\label{fig:haines}
\end{figure}
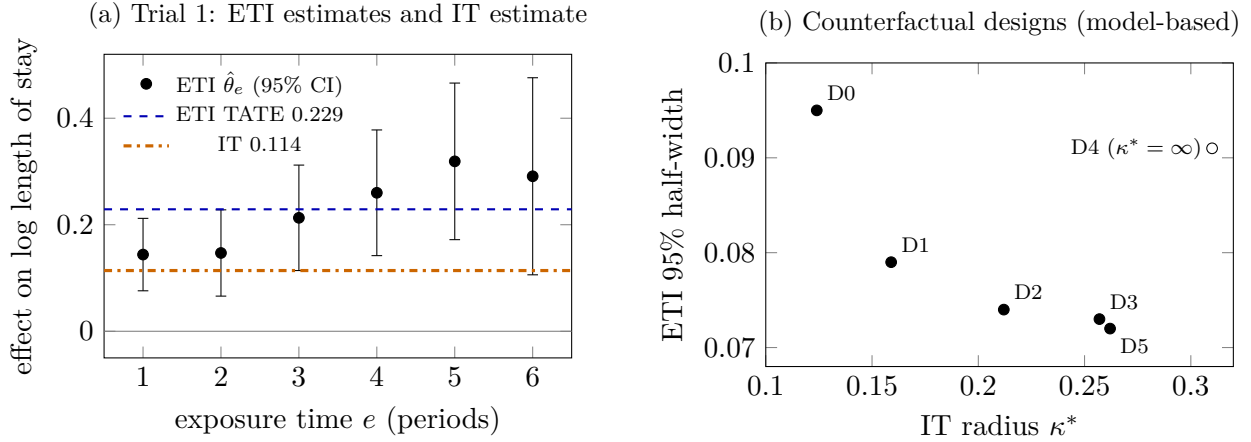

\Cref{tab:haines} and \cref{fig:haines}b give the results. In the trial's design the ETI variance is $2.67$ times the IT variance, within the range reported by \citet{kenny2026factors}. Wait-listing two wards at the same rollout capacity reduces it to $1.86$ times (30\% lower), and four or five wards to $1.60$ or $1.55$ times (40--42\% lower); the ETI 95\% half-width falls from $0.095$ to $0.079$ and $0.073$--$0.074$ on the log scale. Under this model, the ETI analysis of D3 is as precise as the ETI analysis of the trial's complete design would be with about 21 wards instead of 12. The IT variance is unchanged with two wards wait-listed and $7$--$14\%$ higher with four or five. The ETI-optimal design D3 starts six wards in periods 2--4 and one in period 5 and wait-lists five, a front-loaded hybrid like those of \cref{nr:etiopt}. The parallel design with baseline is only slightly more precise than the trial's design for TATE (ETI variance $2.46$ against $2.67$) and much less precise than the hybrids: with $\hat\phi=13.5$ most of the information comes from comparisons within wards over time, which the parallel design uses only through its single baseline period. The small advantage over the stepped wedge agrees with \citet{lee2024cluster}, whose simulated datasets gave standard errors of $0.045$ and $0.048$ for the time-averaged effect in these two designs; our model-based values are $0.047$ and $0.049$. In the simulations the ETI estimator is unbiased (absolute bias at most $0.0024$) and its model-based intervals cover TATE in 94--96\% of trials in every design; the coverage of the IT intervals matches the analytic values.

The hybrid designs do not make the IT analysis safe for this effect curve. Their radii, $0.16$ to $0.26$, remain below the estimated heterogeneity of about $0.64\,\delta_{80}$, and their IT intervals would cover TATE in 14--60\% of trials. What the wait-list buys here is a cheaper robust analysis, which is the practical payoff of \cref{sec:cost}. Calibrating instead to the nested-exchangeable fit gives the same ordering (ETI variances $2.61$, $1.83$, $1.58$, $1.52$, $2.32$ and $1.51$ for D0--D5).

\section{A second trial: digital adherence support for tuberculosis treatment}\label{sec:uganda}

The weekend-services trial shows what happens when heterogeneity is present. The second trial shows the more common situation, in which the data cannot tell. \citet{cattamanchi2021digital} evaluated digital adherence technology (99DOTS) for supervising tuberculosis treatment in 18 health facilities in Uganda, in a stepped wedge with six blocks of three facilities and eight monthly periods. The public file (supplementary file S1 Data of the article) contains 2{,}224 adults initiating treatment for drug-susceptible pulmonary tuberculosis. Each block switched in one of months 1--6, and its switching month was a transition month that the primary analysis excludes, leaving 1{,}022 patients in the control condition and 891 in the intervention condition. Month 0 is all control and month 7 all intervention, so the design is complete; after removal of the transition month, block $b$ is observed at exposures $1,\dots,7-b$ and $L=6$. The outcome is treatment success. \Cref{tab:twotrials} sets the two trials side by side.

\begin{table}[t]
\centering
\begin{threeparttable}
\caption{The two trials side by side. Effects on the log scale of the outcome (log length of stay; log odds of treatment success). $\delta_{80}$ is the effect the trial's IT analysis could detect with 80\% power.}
\label{tab:twotrials}
\footnotesize
\setlength{\tabcolsep}{4pt}
\begin{tabular}{@{}lcc@{}}
\toprule
 & Weekend services (\cref{sec:example}) & TB adherence (\cref{sec:uganda})\\
\midrule
Clusters, periods, outcome & 12 wards, 7, continuous & 18 facilities, 8, binary\\
$\hat\phi$ (working scale) & 13.5 & 0.60\\
IT radius $\kstar$ of the trial's design & 0.124 & 0.122\\
IT estimate; ETI estimate of TATE & 0.114; 0.229 & 0.035; 0.011\\
ETI variance / IT variance (fitted models) & 2.58 & 2.27\\
$\hat\beta_{\mathrm{IT}}-\tauhat$ (SE) & $-0.115$ (0.038) & 0.021 (0.237)\\
Debiased spread of $\hat\theta$, in units of $\delta_{80}$ & 0.64 (5.2 radii) & 0 \\
Spread from estimation error, in units of $\delta_{80}$ & 0.46 & 0.43 (3.5 radii)\\
\addlinespace
\multicolumn{3}{@{}l}{\emph{Counterfactual, same capacity: ETI variance / IT variance of the trial's design}}\\
2 clusters wait-listed & 1.86 & 1.73\\
4 clusters wait-listed & 1.60 & 1.41\\
ETI-optimal number wait-listed & 1.55 (5 of 12) & 1.18 (8 of 18)\\
IT variance at that design & 1.14 & 0.98\\
\bottomrule
\end{tabular}
\begin{tablenotes}\footnotesize
\item Weekend services: random ward intercept, REML. TB adherence: logistic mixed model with facility intercept and covariates, linearised for the design quantities; transition months excluded from the fits and counted as exposure 1 in the counterfactual.
\end{tablenotes}
\end{threeparttable}
\end{table}

\paragraph{Reproduction.} A logistic mixed model with a random facility intercept, categorical month and the covariates that reproduce the published estimate (sex, age, HIV status, disease class, retreatment and facility type), fitted by maximum likelihood with 30-point Gauss--Hermite quadrature, gives an IT odds ratio of $1.04$ ($0.68$ to $1.58$), the published intention-to-treat result to two decimals. The ETI model gives odds ratios of $1.03$, $1.06$, $0.93$, $1.39$, $1.21$ and $0.63$ at exposures $1$--$6$, with log-scale standard errors rising from $0.24$ to $0.62$ (only 35 patients reach exposure 6), and a TATE odds ratio of $1.01$ ($0.53$ to $1.91$), at $2.27$ times the IT variance. The test of a constant effect gives $\chi^2_5=3.9$ ($p=0.56$), and the slope of $\hat\theta_e$ on $e$ is $0.010$ (standard error $0.079$). Leave-one-facility-out jackknife intervals with $t_{17}$ critical values are $0.55$ to $1.97$ for IT and $0.47$ to $2.16$ for TATE (\cref{tab:uganda} in \cref{app:tables}).

\paragraph{Diagnostic.} Linearising the fitted model at its fitted values (15.4 patients per facility-month, random-intercept standard deviation $0.46$) gives $\hat\phi=0.60$ on the working scale. The IT exposure weights are $0.77$, $0.39$, $0.08$, $-0.06$, $-0.11$ and $-0.07$, so $\Gamma=1.91$ and $\kstar=0.122$, the same as in the weekend-services trial; the design with equal cells gives $0.122$ at $\hat\phi$ and $0.103$ under TWFE. The IT analysis is fragile by design in exactly the same way. Here, however, there is no sign that the fragility was exploited: $\hat\beta_{\mathrm{IT}}-\tauhat=(W-u)^\top\hat\theta=0.021$ on the log-odds scale, and the RMS spread of $\hat\theta$ ($0.250$) is no larger than estimation error alone would produce ($0.259$), so the debiased spread is zero.

\paragraph{What the data could have detected.} The trial's IT standard error ($0.216$) implies $\delta_{80}=0.605$ on the log-odds scale (an odds ratio of $1.83$), so the radius corresponds to an RMS spread of $0.074$. The standard error of $(W-u)^\top\hat\theta$ is $0.237$, larger than the IT standard error itself, and the spread produced by estimation error alone is $0.43\,\delta_{80}$, three and a half times the radius. Against the least-favourable profile $\theta-\tau\one\propto W-u$, the 5-degree-of-freedom test of a constant effect has power $8\%$, $18\%$, $37\%$ and $85\%$ when the spread is one, two, three and five times the radius; at three times the radius the IT bias is $1.96$ standard errors and the nominal IT interval covers TATE in $50\%$ of trials. A null result for heterogeneity in this trial is therefore uninformative about whether the IT analysis is adequate. The published conclusion is probably not distorted, since the estimated curve gives no reason to think so, but the design could not have shown that it was not.

\paragraph{Counterfactual redesign.} As in \cref{sec:example}, outcomes under the alternative designs were not observed. Using the linearised working model with $\hat\phi=0.60$ and equal cells, counting the transition month as the first exposure month (as in the trial's prespecified sensitivity analysis), and keeping the trial's capacity of three facilities switching per month, the ETI variance relative to the IT variance of the trial's design is $2.54$ with no facility wait-listed, $1.73$ with two, $1.41$ with four, $1.25$ with six and $1.18$ with eight, the minimum. Unlike in the weekend-services trial, the IT variance does not rise: it is $0.88$--$0.98$ times that of the trial's design for two to eight wait-listed facilities, and the IT radius rises from $0.128$ to $0.162$, $0.198$, $0.254$ and $0.366$. With a weak within-facility correlation, much of the information comes from comparisons between facilities in the same month, which a wait-list adds. In this trial a wait-list would have made the robust analysis affordable at no cost to the conventional one.

\section{Discussion}\label{sec:discussion}

\subsection{Implications for design and analysis plans}

The results lead to five recommendations for trials in which the effect may build up or wane after implementation.

First, a protocol that plans an IT analysis of a complete stepped wedge should state that the design tolerates exposure-time variation of only about 12--16\% of the planned effect at realistic intracluster correlations (\cref{thm:twfe,thm:gls}), and that under a linear learning curve the fixed-effects IT estimate refers to the effect half a period after implementation (\cref{prop:ramp}). These numbers require only the number of periods and a planning value of $\phi$.

Second, more clusters per step, reallocation across steps and periods added after the rollout do not change this materially (\cref{thm:twfe}, \cref{nr:realloc}, \cref{cor:final}). Extra baseline periods help modestly.

Third, if heterogeneity is plausible, the analysis plan should target TATE directly with an ETI model, or declare a heterogeneity level $\kappa$ and report a bias-aware IT interval when $\kappa$ is below the crossover of \cref{prop:crossover}, which is about $0.27$ for common designs.

Fourth, the design should then be chosen for the analysis it will receive. Keeping a wait-listed arm under the same rollout capacity reduces the cost of the ETI analysis sharply, and most of the reduction comes from the first few wait-listed clusters (\cref{nr:frontier}); in our settings, wait-listing a third of clusters did not increase the IT variance either (\cref{tab:levers}). Wait-listed clusters can receive the intervention after follow-up, so the design keeps the feature that every cluster is eventually treated.

Fifth, a wait-list also protects against heterogeneity that ETI does not address. Without concurrent controls in the final period, calendar-time and cohort-specific effects there are not identified (\cref{prop:identification}); with them, they are.

The weekend-services trial shows these points in one dataset. Its design had the radius predicted by \cref{thm:gls}, its IT estimate was about half of the ETI estimate of TATE, the data could not separate exposure-time from calendar-time variation, and, under a model calibrated to it, a modest wait-list would have made the ETI analysis substantially more precise without changing the rollout capacity (\cref{sec:example}). Reporting $\hat\beta_{\mathrm{IT}}-\tauhat=(W-u)^\top\hat\theta$ with its standard error, as in \cref{sec:example}, is a simple way to show how much an IT analysis depends on the absence of exposure-time variation. The tuberculosis adherence trial (\cref{sec:uganda}) shows why the standard error matters as much as the estimate. Its design had the same radius, its estimated contrast was close to zero, and its test of a constant effect found nothing, but that test would have missed heterogeneity large enough to halve the coverage of the IT interval nearly two times in three. In a complete design, a null check of this kind does not show that the IT analysis is adequate; a design with concurrent controls makes the check informative at an affordable cost.

\subsection{Relation to existing work}

The heterogeneity problem and the ETI remedy are due to \citet{kenny2022analysis}, and the precision cost of the remedy was quantified by \citet{kenny2026factors}, who identified design for time-varying estimands as open. Our design results answer part of that question for TATE. Design under exposure-time heterogeneity is therefore not new: \citet{lee2024cluster} compared the variances of exposure-time estimators in four given designs, showed that the parallel design with baseline is usually the most efficient of them for time-averaged and long-term effects and that its IT and time-averaged estimators coincide, and calibrated simulations to the same trial we use. Our contribution differs in criterion, explanation and search space. The adequacy radius measures how the misspecified IT analysis behaves rather than how precise the correctly specified one is. \Cref{lem:balance} and \cref{prop:identification} explain the ranking of their designs: the stepped wedge has no concurrent controls in its final period, and the designs that add them do better. And we optimise over intermediate allocations under a rollout capacity, which places the parallel design with baseline at one end of a family and shows that front-loaded hybrids between it and the stepped wedge are at least as precise as either in all the settings we studied, and often clearly more precise. Hybrid designs are precision-optimal under a constant effect \citep{girling2016statistical,thompson2017optimal}; under exposure-time heterogeneity the same design feature, concurrent controls in every period, matters much more, and in our computations the ETI-optimal hybrids wait-list about twice as many clusters as the constant-effect optimum (\cref{sec:cost}). The optimal-design machinery reviewed by \citet{watson2023optimal} applies directly to the ETI criterion, since minimising $\Var(\tauhat)$ is a $c$-optimality problem; what changes is the contrast vector. The non-identification of final-period effects without never-treated units is known in the difference-in-differences literature \citep{borusyak2024revisiting}; \cref{lem:balance} is its elementary counterpart for linear estimators, and in that form it also accounts for the weights of the IT estimator and the cost of the ETI estimator.

\subsection{Limitations}

The theory assumes balanced panels with equal cluster-period sizes, cross-sectional sampling, a continuous outcome and the exchangeable correlation of \citet{hussey2007design}. Block-exchangeable and decaying correlation, a binary outcome on the linearised logit scale and a misspecified working model are covered numerically only (\cref{nr:etiopt}); cohort designs and unequal cluster sizes are not yet covered. The radius uses the true standard error and Gaussian planning; finite-sample behaviour of feasible intervals has been checked only in one setting (\cref{sec:smallG}). The target is the equal-weight TATE over exposures $1,\dots,T-1$, and therefore depends on the design horizon, as \citet{kenny2022analysis} note. The heterogeneity classes concern exposure time only; calendar-time heterogeneity enters through \cref{prop:identification} and remains a sensitivity class rather than part of the bounds. The reallocation ceiling (\cref{conj:ceiling}), hybrid allocations outside the balanced family of \cref{thm:hybridgls}, and all statements about ETI cost and ETI-optimal designs are numerical. A wait-list withholds the intervention from some clusters for the whole follow-up. That is the same requirement as in a parallel design, but it may be unacceptable where the stepped wedge was chosen precisely to avoid it. In the illustration, time is indexed by study period because calendar dates are released for only one hospital; the redesign exercise assumes equal ward-period sizes, a random-intercept covariance and the trial's estimated effect curve as truth, and its results are model-based and simulated, not observed. Information criteria computed on the same data show only a moderate preference for a calendar-time effect, which the complete design cannot confirm or exclude. In the tuberculosis adherence trial the design quantities come from linearising a logistic mixed model with covariates, so the radius, power and counterfactual variances are approximations on the log-odds scale; the counterfactual counts the excluded transition month as the first exposure month; and the intervention supports taking a medication rather than being one.

\subsection{Scope, interpretation, and remaining generality}
The adequacy radius is conditional on a target estimand, a heterogeneity class, and a tolerance; it is not a universal certificate that the constant-effect model is scientifically adequate. The analytic monotonicity results (\cref{thm:hybridgls} for TATE, Appendix~\ref{app:balanced-waitlist-monotonicity} for the design-average exposure target) use balanced sequence allocation, equal cluster-period information and compound-symmetric covariance. It does not establish monotonic ETI precision under arbitrary wait-listing, nor does it characterise the globally optimal hybrid design. The corresponding optimisation results should therefore be read as reproducible computations within the stated design class, not as theorems.

The Haines redesign is likewise a counterfactual, model-based calculation. The observed trial and its outcomes are real, but outcomes under alternative hybrid allocations were not observed. Its purpose is to calibrate the diagnostic and show the scale of the design trade-off, not to claim that the historical investigators should have chosen a particular alternative. Our calendar-time comparisons refit candidate specifications to the public data under a common coding and likelihood convention. They should not be interpreted as demonstrating an error in earlier work: differences can arise from outcome construction, missing-data handling, time coding, fixed- versus random-effect choices, or likelihood conventions.

Finally, the current calculations do not exhaust unequal cluster sizes, non-Gaussian outcomes, small-cluster finite-sample corrections, alternative within-cluster correlation structures, or ethical and logistical restrictions on maintaining a wait-list. These features can alter both the feasible design set and the relevant precision criterion. They are important extensions, but adding a broad simulation catalogue here would obscure the paper's sharper design principle. The practical recommendation is therefore diagnostic: specify the intended exposure-time target, examine the constant-effect estimand against a plausible heterogeneity class, and retain enough concurrent control information to make a target-aligned analysis viable when the diagnostic indicates material risk.

\subsection{Conclusion}

A complete stepped wedge analysed with an IT model tolerates little exposure-time heterogeneity, and almost the same amount whatever its size, because its final periods contain no concurrent controls. Restoring concurrent controls with a wait-listed arm raises that tolerance and, more usefully, makes the robust ETI analysis much cheaper. When the effect of an intervention may change over time, the choice between a complete wedge and a hybrid should be made together with the choice of analysis.

\section*{Code and data availability}
Code reproducing every numerical result, together with the outputs it produced, is available from the authors on request and will be deposited in a public repository.
The trial data used in \cref{sec:example} are publicly available as supplementary file S2 of \citet{haines2017impact}; the analysis is \texttt{haines\_trial1\_analysis.py}. The data used in \cref{sec:uganda} are publicly available as supplementary file S1 Data of \citet{cattamanchi2021digital} (figshare item 14550544); the analysis is \texttt{uganda\_reanalysis.py} with \texttt{uganda\_reanalysis\_extra.py}.

\bibliographystyle{plainnat}
\bibliography{references}

\appendix
\section{Balanced wait-list monotonicity under compound symmetry}
\label{app:balanced-waitlist-monotonicity}

This section isolates the mechanism behind the computational wait-list pattern in a deliberately narrow family. The result is not a claim about arbitrary hybrid allocations, unequal cluster sizes, alternative correlation structures, or the variance of the ETI estimator. It concerns the implicit estimand of the period-adjusted constant-effect GLS estimator in a balanced Hussey--Hughes calculation.

Consider $J=S+1$ periods and $S\geq 2$ switching sequences. Sequence $s\in\{1,\ldots,S\}$ is untreated through period $s$ and treated thereafter. A proportion $a/S$ of clusters is assigned to each switching sequence and a proportion $1-a$ remains untreated throughout follow-up, where $0<a\leq 1$. Thus $1-a$ is the wait-list share among all clusters. At the cluster-period-mean level, let the covariance be compound symmetric with correlation $\rho\in[0,1)$ and write
\[
  \gamma=\frac{\rho}{1+S\rho},
\]
so that the inverse covariance is proportional to $I-\gamma\mathbf 1\mathbf 1^{\mathsf T}$. Let $\delta_r$ denote the effect at exposure time $r\in\{1,\ldots,S\}$, and let $w_r(a,\rho)$ be the implicit exposure-time weights of the constant-effect GLS coefficient after adjustment for unrestricted period effects.

For the design-average target, exposure time $r$ occurs in $n_r=S-r+1$ rollout sequence-periods. Hence
\[
  q_r=\frac{n_r}{Sm},\qquad m=\frac{S+1}{2}.
\]
Also define
\[
 b_2=\frac{(S+1)(2S+1)}{6},
 \qquad h_r=\frac{S+r}{2},
\]
and
\[
 H(a,\gamma)=m-\gamma b_2-a\left(\frac{b_2}{S}-\gamma m^2\right).
\]
Direct residualisation of the treatment indicator with respect to period effects gives
\begin{equation}
\label{eq:balanced-waitlist-weight}
 w_r(a,\rho)
 =\frac{(n_r/S)\{1-\gamma h_r-a(h_r/S-\gamma m)\}}
        {H(a,\gamma)}.
\end{equation}
The weights sum to one. Subtracting the design-average target yields the factorisation
\begin{equation}
\label{eq:balanced-waitlist-factorisation}
 w_r(a,\rho)-q_r
 = C(a,\gamma)\,\frac{n_r}{S}\{b_2-mh_r\},
 \qquad
 C(a,\gamma)=\frac{\gamma+a/S}{mH(a,\gamma)}.
\end{equation}
Thus the entire mismatch vector changes only through the positive scalar $C(a,\gamma)$.

\paragraph{Proposition (balanced wait-list monotonicity).}
For $S\geq2$, $0<a\leq1$, and $0\leq\rho<1$, every positively homogeneous norm of $w(a,\rho)-q$ is strictly increasing in the rollout share $a$ and strictly increasing in $\rho$. Equivalently, the mismatch strictly decreases as the wait-list share $1-a$ rises, and increases as within-cluster correlation rises. Any adequacy radius obtained by dividing a fixed tolerated estimand error by such a mismatch norm therefore moves in the opposite direction.

\paragraph{Proof.}
Let $A=m-\gamma b_2$ and $B=b_2/S-\gamma m^2$, so that $H=A-aB$. The scalar in Equation~\eqref{eq:balanced-waitlist-factorisation} differs only by the constant $1/m$ from
\[
 F(a,\gamma)=\frac{\gamma+a/S}{A-aB}.
\]
Differentiation and simplification give
\[
 \frac{\partial F}{\partial a}
 =\frac{m\{1/S-\gamma^2m\}}{H(a,\gamma)^2}>0,
 \qquad
 \frac{\partial F}{\partial\gamma}
 =\frac{m\{1-a^2m/S\}}{H(a,\gamma)^2}>0.
\]
The first inequality follows from $\gamma<1/(S+1)$; the second follows from $m/S=(S+1)/(2S)<1$ for $S\geq2$. Moreover, $\gamma$ is strictly increasing in $\rho$. The vector multiplying $C(a,\gamma)$ in Equation~\eqref{eq:balanced-waitlist-factorisation} is nonzero for $S\geq2$. Positive homogeneity therefore establishes both strict monotonicity statements. \hfill$\square$

The limiting calculation also exposes a reallocation ceiling. As the rollout share tends to zero,
\[
 C(a,\gamma)\longrightarrow
 \frac{\gamma}{m(m-\gamma b_2)}.
\]
The mismatch therefore vanishes in this limit only when $\rho=0$; positive within-cluster correlation leaves a nonzero floor. Relative to the complete wedge ($a=1$), the limiting mismatch ratio is
\begin{equation}
\label{eq:balanced-waitlist-floor-ratio}
 \frac{\gamma(S-1)(2-\gamma S)}
 {2(1+\gamma S)\{3-\gamma(2S+1)\}}.
\end{equation}
At the same time, treatment information is proportional to $aH(a,\gamma)$ and tends to zero with $a$. The limit is therefore a structural boundary, not a design recommendation: retaining concurrent controls improves estimand alignment, but reallocating nearly all clusters to control destroys precision. This is the analytic core of the robustness--precision frontier.

\section{Proofs}\label{app:proofs}

Throughout, $\lVert\cdot\rVert$ is the Euclidean norm and $u=L^{-1}\one$. Every closed form below was also verified in exact rational arithmetic against weights computed by solving the GLS or TWFE normal equations, for $T=3,\dots,10$ (wider ranges for the bounds) and several rational values of $\phi$ and $q$, by the script \texttt{verify\_closed\_forms\_exact.py}, whose output also lists every closed-form value quoted in the text. \Cref{thm:hybridgls} is checked separately: symbolically for general $T$, $q$ and $\phi$ by \texttt{proof\_tate\_hybrid.py}, and, in exact rational arithmetic and against full GLS fits, for $T=3,\dots,10$ by \texttt{proof\_tate\_hybrid\_check.py}.

\paragraph{Proof of \cref{prop:worstbias}.} By \eqref{eq:model} and period invariance the bias is $(W-u)^\top\theta$. Because $\one^\top(W-u)=0$, it equals $(W-u)^\top(\theta-\tau\one)$. By definition $\lVert\theta-\tau\one\rVert=\sqrt L\,s(\theta)$, so by the Cauchy--Schwarz inequality the bias is at most $\lVert W-u\rVert\sqrt L\,s=\Gamma s$ in absolute value. Equality holds for $\theta-\tau\one\propto W-u$, which is admissible because $W-u$ sums to zero, so $\tau$ is unchanged. \qed

\paragraph{Proof of \cref{lem:balance} and \cref{cor:final}.} Under \eqref{eq:model}, $\E\betahat=\sum_t\mu_t\sum_ca_{ct}+\sum_eW_e\theta_e$ (plus $\sum_c\alpha_c\sum_ta_{ct}$ when cluster effects are fixed). Invariance for all $(\mu_t)$ requires $\sum_ca_{ct}=0$. If all clusters are treated in period $t$, $\sum_ca_{ct}D_{g(c)t}=\sum_ca_{ct}=0$. In a complete design every cluster is treated in period $T$; the treated cells of period $T$ have exposures $T-g+1$, $g=2,\dots,T$, and exposure $L$ occurs only for $g=2$, which gives the corollary. \qed

\paragraph{Proof of \cref{prop:identification}.} (a) Take expectations in \eqref{eq:model} conditional on the cluster effects; the period effect cancels. (b) In a complete design the means of period $T$ are $\mu_T+\bar\alpha_g+\tau_{gT}$ for every $g$, and $\tau_{gT}$ appears nowhere else. The map $(\mu_T,\tau_{\cdot T})\mapsto(\mu_T+c,\tau_{\cdot T}-c)$ leaves the distribution of the data unchanged for every $c$, so only differences $\tau_{gT}-\tau_{g'T}$ are identified, and a positively weighted average shifts by $-c\sum w_g\neq0$. If $p_\infty>0$, the wait-listed means satisfy $\E(\bar Y_{\infty t}-\bar Y_{\infty1})=\mu_t-\mu_1$, and since period 1 is untreated in every sequence, $\E(\bar Y_{gt}-\bar Y_{g1})=\mu_t-\mu_1+\tau_{gt}$ for $t\ge g$. The difference identifies $\tau_{gt}$. \qed

\paragraph{Proof of \cref{thm:twfe}.} Replicating every sequence leaves all ratios unchanged, so take one cluster per sequence and index it by $i=g-1\in\{1,\dots,T-1\}$. In a balanced panel the two-way residual of $D$ is exact: $\tilde D_{it}=D_{it}-\bar D_{i\cdot}-\bar D_{\cdot t}+\bar D$, with $\bar D_{i\cdot}=(T-i)/T$, $\bar D_{\cdot t}=(t-1)/(T-1)$ and $\bar D=1/2$. By the Frisch--Waugh--Lovell theorem $\betahat_W=\sum\tilde D_{it}\bar Y_{it}/n_w$ with $n_w=\sum\tilde D_{it}^2=\sum_{\text{treated}}\tilde D_{it}$. The treated cells with exposure $e$ are $(i,i+e)$, $i=1,\dots,T-e$, on which $\tilde D$ is linear in $i$; summing gives
\[
N_e=\sum_{i=1}^{T-e}\tilde D_{i,i+e}=\frac{(T-e)\{(T^2-1)-(2T-1)e\}}{2T(T-1)},\qquad n_w=\sum_eN_e=\frac{(T-2)(T+1)}{12},
\]
and $\Wtw_e=N_e/n_w$ is \eqref{eq:Wtwfe}. Expanding $\Gamma^2=L\sum_e(\Wtw_e-1/L)^2$ with Faulhaber's formulas gives \eqref{eq:G2twfe}. The values of $\Wtw_1$, $\Wtw_L$ and the sign condition are read off \eqref{eq:Wtwfe}. The identity for $19/5-\Gamma^2$ is direct, and $T^2-3T-1>0$ for $T\ge4$. The derivative $\partial\Gamma^2/\partial T$ has the sign of $T^4-6T^3+2T^2+2T+2$, which is negative at $T=4,5$ and positive for $T\ge6$, so over integers $T\ge4$ the minimum of $\Gamma^2$ is at $T=6$, where it equals $249/70$. With $\kstar=0.23285/\Gamma$ this gives the bounds. At $T=3$, $\Gamma^2=4$. \qed

\paragraph{Proof of \cref{lem:wb}.} Let $Y_c,X_c$ be the $T$-vectors (or $T\times k$ matrices) of cluster $c$. GLS minimises $\sum_c(Y_c-\mu-X_c\beta)^\top V^{-1}(Y_c-\mu-X_c\beta)$. The normal equations for $\mu$ are $V^{-1}\sum_c(Y_c-\mu-X_c\beta)=0$; since $V^{-1}$ is invertible and common to all clusters, $\hat\mu=\bar Y-\bar X\hat\beta$, where bars denote averages over clusters. Substituting, $\hat\beta$ minimises $\sum_c(\tilde Y_c-\tilde X_c\beta)^\top V^{-1}(\tilde Y_c-\tilde X_c\beta)$ with period-centred data. Write $V^{-1}=Q+\lambda P$ with $Q=I-J/T$, $P=J/T$ and $\lambda=1/(1+T\phi)$. Then $\hat\beta=(S_W+\lambda S_B)^{-1}(s_W+\lambda s_B)$ with $S_W=\sum\tilde X_c^\top Q\tilde X_c$, $s_W=\sum\tilde X_c^\top Q\tilde Y_c$, $S_B=\sum\tilde X_c^\top P\tilde X_c$, $s_B=\sum\tilde X_c^\top P\tilde Y_c$. The within part is TWFE, because $Q$ removes cluster means and period-centring removes period effects; the between part is least squares on period-centred cluster means. For scalar $\beta$, $S_W=\sum\tilde D_{ct}^2$ and $S_B=T\sum_c(\bar D_c-\bar D)^2$, which gives (ii). Both component estimators have exposure weights summing to one, so the weights combine with the same $\omega$. Nothing in the argument uses $k=1$, which gives (iii). \qed

\paragraph{Proof of \cref{thm:gls}.} (i) With one cluster per sequence, the period-centred cluster means of $D$ are $\bar D_{i\cdot}-\bar D=(T-2i)/(2T)$. The between estimator puts weight proportional to $\bar D_{i\cdot}-\bar D$ on every cell of cluster $i$; these weights already sum to zero within each period. Summing over the treated cells with exposure $e$ gives $\sum_{i=1}^{T-e}(T-2i)/(2T)=(T-e)(e-1)/(2T)$, and normalising gives $\Wbt_e$. Also $S_B=T\sum_i\{(T-2i)/(2T)\}^2=(T-2)(T-1)/12$, and $S_W=n_w=(T-2)(T+1)/12$ from the proof of \cref{thm:twfe}, so $\omega=(T+1)/\{(T+1)+(T-1)\lambda\}$. The identity with equation~(5) of \citet{kenny2022analysis} under $\psi=\phi/(1+\phi)$ and $Q=T-1$ was checked in exact arithmetic.

(ii) Let $a=\Wtw-u$ and $b=\Wbt-u$, and $A=L\lVert a\rVert^2$, $B=L\lVert b\rVert^2$, $C=La^\top b$. Then $A$ is \eqref{eq:G2twfe}; using $\sum_{j=0}^{N}j^2(N-j)^2=N(N^4-1)/30$, $B=(T^2-2T+12)/\{5T(T-2)\}$; and $C=-2(T^3+5T^2-2T-6)/\{5T(T-2)(T+1)\}$. Substituting $\omega(\phi)$ into $\Gamma^2=A\omega^2+2C\omega(1-\omega)+B(1-\omega)^2$ gives \eqref{eq:G2gls}.

(iii) $f(\omega)=\lVert\omega a+(1-\omega)b\rVert^2$ is convex. At $\omega_0=(T+1)/(2T)$, the value at $\phi=0$, $(\omega_0a+(1-\omega_0)b)^\top(a-b)=3(3T-4)/\{5(T-2)(T-1)\}>0$, so $f'(\omega_0)>0$, and by convexity $f$ is strictly increasing on $[\omega_0,1]$. Since $\omega(\phi)$ is strictly increasing in $\phi$ with $\omega(0)=\omega_0$, $\Gamma$ is strictly increasing in $\phi$.

(iv) Setting $\phi=0$ in \eqref{eq:G2gls} gives $(2T-1)(2T-3)/\{5T(T-2)\}$, which exceeds $4/5$ because $(2T-1)(2T-3)-4T(T-2)=3$. Monotonicity in $\phi$ and \cref{thm:twfe} give $4/5<\Gamma^2<19/5$ for $T\ge4$, and the bounds on $\kstar$ follow. \qed

\paragraph{Proof of \cref{prop:ramp}.} (i) $\sum_e e\Wtw_e=1/2$ for every $T$ by Faulhaber's formulas. $\Wbt$ is symmetric under $e\mapsto T+1-e$ and sums to one, so $\sum_ee\Wbt_e=(T+1)/2$. Hence $\ebar=\omega/2+(1-\omega)(T+1)/2$. At $\phi=0$, $\omega_0=(T+1)/(2T)$ gives $\ebar=(T+1)/4$, and $\ebar$ decreases in $\omega$. For $\theta_e=\theta_1+b(e-1)$, $\E\betahat=\sum_eW_e\theta_e=\theta_1+b(\ebar-1)$, and $\tau=\theta_1+b(T/2-1)$, so the bias is $b(\ebar-T/2)=-\tfrac b2(\omega T-1)$.

(ii) Write $d=W-u$, $S_k=\sum_{e\le k}d_e$ and $R_k=\sum_{e>k}d_e=-S_k$. Summation by parts gives $(W-u)^\top\theta=\sum_{k=1}^{L-1}R_k(\theta_{k+1}-\theta_k)$, so $\sup_{\Theta_M}|\text{bias}|=M\sum_k|R_k|$, attained by $\theta_{k+1}-\theta_k=M\,\mathrm{sign}(R_k)$. We show $S_k>0$ for $k<L$. Under TWFE, $\Wtw_e$ is a convex quadratic in $e$ with $\Wtw_1=6/(T+1)>1/L$ and $\Wtw_L<0<1/L$; the set where a convex function lies below $1/L$ is an interval, which here contains $L$ but not $1$, so $d_e$ changes sign once, from positive to negative. Hence $S_k$ rises and then falls to $S_L=0$, and $S_k>0$ for $k<L$. At $\phi=0$, $W_e(0)=3(T-e)(T-1-e)/\{T(T-1)(T-2)\}$ is strictly decreasing in $e$, so the same argument applies. For general $\phi$, $S_k(\phi)$ is affine in $\omega\in[\omega_0,1]$ and positive at both ends, hence positive. All $R_k$ are therefore negative, the extremal profile is a linear ramp, and $M\sum_k|R_k|=M\sum_k(-R_k)=M\,(T/2-\ebar)=\tfrac M2(\omega T-1)$.

(iii) A ramp with slope $b$ has $s=|b|\{(L^2-1)/12\}^{1/2}$. Under TWFE its bias is $|b|L/2$, so it attains $\Gamma_{\mathrm{ramp}}=\sqrt3L/(L^2-1)^{1/2}$ against $\Gamma$ of \eqref{eq:G2twfe}. The stated values follow; as $T\to\infty$, $\Gamma_{\mathrm{ramp}}\to\sqrt3$ and $\Gamma\to(19/5)^{1/2}$. \qed

\paragraph{Proof of \cref{prop:t3}.} The exact two-way residuals weighted by the shares are $r_A=\tfrac{1-p}3(-1,2,-1)$ for the sequence starting in period 2 and $r_B=\tfrac p3(1,-2,1)$ for the other, so $n_w=2p(1-p)/3$. Then $W_1=\{p\cdot2(1-p)/3+(1-p)p/3\}/n_w=3/2$ and $W_2=-\{p(1-p)/3\}/n_w=-1/2$. \qed

\paragraph{Proof of \cref{prop:parallel}.} For a balanced panel with cluster shares $p_c$, the weighted two-way residual $r_{ct}=D_{ct}-\bar D_c-P_t+\bar D$, with $P_t=\sum_cp_cD_{ct}$, sums to zero over $t$ and, with weights $p_c$, over $c$; hence it is the exact residual. In $\mathcal{P}_T(p)$, $\bar D_c=(T-1)/T$ for treated clusters and $0$ for wait-listed ones, $P_1=0$, $P_t=p$ for $t\ge2$, and $\bar D=p(T-1)/T$. Every treated cell therefore has $r=1-(T-1)/T-p+p(T-1)/T=(1-p)/T$, the same value, and each exposure occurs in exactly one treated cell per treated cluster. So $\Wtw$ is uniform. The between estimator gives every treated cell a weight proportional to $\bar D_c-\bar D$, again constant, so $\Wbt$ is uniform. By \cref{lem:wb}, $W(\phi)$ is uniform for every $\phi$. \qed

\paragraph{Proof of \cref{thm:hybrid}.} (i) Use the weighted residual of the previous proof. In $\mathcal{H}_T(q)$, with stepped sequences indexed by $i=g-1$, $\bar D_i=(T-i)/T$, $\bar D_\infty=0$, $P_t=(1-q)(t-1)/(T-1)$ and $\bar D=(1-q)/2$. On the treated cell $(i,i+e)$, $r=1-(T-i)/T-(1-q)(i+e-1)/(T-1)+(1-q)/2$. Summing over $i=1,\dots,T-e$ with weight $(1-q)/(T-1)$ gives
\[
N_e=\frac{(1-q)(T-e)\{(T^2-1)-(2T-1)e+Tqe\}}{2T(T-1)^2},\qquad n_w=\frac{(1-q)(T+1)(T-2+Tq)}{12(T-1)},
\]
and $W_e(q)=N_e/n_w$. Splitting the numerator as $6(T-e)\{(T^2-1)-(2T-1)e\}+6Tq\,e(T-e)$ gives the mixture with $\pi_q=Tq/(T-2+Tq)$, using $\sum_{e=1}^{T-1}e(T-e)=T(T-1)(T+1)/6$.
(ii) $\Gamma^2$ follows from Faulhaber's formulas; it reduces to \eqref{eq:G2twfe} at $q=0$. Differentiating gives the stated derivative. For $q\in[0,1]$ and $T\ge3$, $q(T^2-3T)\le T^2-3T<7T^2-7T-2$, so the derivative is negative.
(iii) $W_e(q)<0$ if and only if $e>(T^2-1)/(2T-1-Tq)$. At $q=(T-2)/T$ the threshold equals $L$, and it increases with $q$.
(iv) $V$ is symmetric about $T/2$ and sums to one, so $\sum_eeV_e=T/2$, and $\ebar(q)=(1-\pi_q)/2+\pi_qT/2$. The linear-ramp bias is $b\{\ebar(q)-T/2\}=-(1-\pi_q)bL/2$.
(v) Setting $q=1$ gives $(T-2)(4T+3)/\{5T(T+1)\}$, which increases in $T$ from $0.38$ at $T=4$ to $4/5$; the radius bounds follow. \qed

\paragraph{Proof of \cref{thm:hybridgls}.}
(i) By \cref{lem:wb}, which applies because wait-listed clusters are observed in every period, $W=\omega\Wtw+(1-\omega)\Wbt$ with $\omega=S_W/(S_W+\lambda S_B)$. By \cref{thm:hybrid}(i), $\Wtw_e(q)\propto(T-e)\{(T^2-1)-(2T-1-Tq)e\}$, and from its proof $S_W=n_w=(1-q)(T+1)(T-2+Tq)/\{12(T-1)\}$. For the between estimator, with stepped sequences indexed by $i=g-1$, $\bar D_i-\bar D=(T-i)/T-(1-q)/2$ and wait-listed clusters have $\bar D_\infty-\bar D=-(1-q)/2$. Hence
\[
S_B=T\Big\{\frac{1-q}{T-1}\sum_{i=1}^{T-1}(\bar D_i-\bar D)^2+q\frac{(1-q)^2}{4}\Big\}=\frac{(1-q)(T-2+3Tq)}{12},
\]
and summing the cell weights $(\bar D_i-\bar D)/S_B$, with share $(1-q)/(T-1)$, over the treated cells $(i,i+e)$, $i=1,\dots,T-e$, gives
\[
\Wbt_e(q)=\frac{6(T-e)(e-1+Tq)}{T(T-1)(T-2+3Tq)} .
\]
Both $\Wtw$ and $\Wbt$ have the form $(T-e)\ell(e)$ with $\ell$ affine, and so does any combination of them. Since $\sum_e(T-e)=T(T-1)/2$ and $\sum_ee(T-e)=T(T-1)(T+1)/6$, a vector $(T-e)(1-te)$ normalised to sum to one has the stated form. Substituting $\omega$, $\lambda=1/(1+T\phi)$, $\Wtw$ and $\Wbt$ and collecting terms gives $t(q,\phi)$; at $\phi=\infty$, $\omega=1$ and $t$ is read off $\Wtw$. Since $t$ is increasing in $\phi$ by (iii), its range is $[t(q,0),t(q,\infty)]$, and $t(q,\infty)\le(2T-1)/(T^2-1)<3/(T+1)$; $t(q,0)=(1-q)/\{(T-1)(1+q)\}\ge0$.

(ii) With $W_e=(T-e)(1-te)/N(t)$, $N(t)=T(T-1)\{3-(T+1)t\}/6$, expanding $\Gamma^2=L\sum_e(W_e-1/L)^2$ with Faulhaber's formulas gives the stated expression. Its derivative is
\[
\frac{d\Gamma^2}{dt}=\frac{6(T-2)(T+1)\{(T-3)t+5\}}{5T\{3-(T+1)t\}^3},
\]
which is positive for $t\in[0,3/(T+1))$ and $T\ge3$ (at $T=3$, $\Gamma^2=1/(3-4t)^2$).

(iii) Differentiating,
\[
\frac{\partial t}{\partial q}=-\frac{T(T+1)\phi^2+4T\phi+2}{(T-1)\{(1+q)+(T+1)\phi\}^2}<0,\qquad
\frac{\partial t}{\partial \phi}=\frac{T(1+2q-q^2)-2}{(T-1)\{(1+q)+(T+1)\phi\}^2}>0,
\]
the latter because $1+2q-q^2\ge1$ on $[0,1]$ and $T\ge3$. At $\phi=\infty$, $t(q,\infty)$ is strictly decreasing in $q$ directly, and $t(q,\phi)\uparrow t(q,\infty)$. Combined with (ii), $\Gamma$ is strictly decreasing in $q$ and strictly increasing in $\phi$, and $\kstar=0.2329/\Gamma$ moves in the opposite directions.

(iv) $t$ is minimised over the family at $\phi=0$ and $q\to1$, where $t\to0$, and $\Gamma^2(0)=15(T-2)/(45T)=(T-2)/(3T)$. Since $q<1$, $t>0$ and the inequality is strict.

(v) $\sum_ee(T-e)(1-te)=T(T-1)(T+1)\{2-Tt\}/12$ by Faulhaber's formulas; dividing by $N(t)$ gives $\ebar$. For $\theta_e=\theta_1+b(e-1)$ the bias is $b(\ebar-T/2)=-b(T-2)/[2\{3-(T+1)t\}]$, whose absolute value increases with $t$. \qed

\paragraph{Proof of \cref{prop:frontier}.} The feasible set at level $\gamma$ is convex, because the constraints are linear except $\sqrt L\lVert W-u\rVert\le\gamma$, and $W$ is linear in the estimator's coefficients. If $a_1$ is feasible at $\gamma_1$ and $a_2$ at $\gamma_2$, then $ta_1+(1-t)a_2$ is feasible at $t\gamma_1+(1-t)\gamma_2$ by the triangle inequality, and the variance, a convex quadratic, is at most $tF_d(\gamma_1)+(1-t)F_d(\gamma_2)$ at the optimum; hence $F_d$ is convex. It is non-increasing because the feasible sets grow with $\gamma$. At $\gamma=0$ the constraints say that the estimator is unbiased for $\tau$ under \eqref{eq:model} for every $(\mu,\theta)$, and the minimum-variance such estimator is the GLS estimator of $\tau$ in the ETI model by the Gauss--Markov theorem. Dropping the constraint on $\Gamma$ leaves the estimators unbiased under a constant effect, among which IT-GLS has minimum variance; it is feasible once $\gamma\ge\Gamma(\betahat_{\mathrm{IT\text{-}GLS}})$. \qed

\paragraph{Proof of \cref{prop:crossover}.} The bias-aware IT half-length is $\mathrm{cv}_\alpha(\bar b/\SD)\SD$ with $\bar b=\Gamma\kappa\delta$ by \cref{prop:worstbias}, and $\delta=2.8016\,\SD$ at planned power. The ETI half-length is $1.96\,\varrho\,\SD$. Dividing by $\SD$ gives the condition; $\mathrm{cv}_\alpha$ is continuous and strictly increasing in its argument, which gives a unique crossover. \qed

\section{Supplementary tables}\label{app:tables}

\begin{table}[h]
\centering
\caption{Radius $\kappa^*$ of the IT analysis of the balanced complete wedge at 80\% planned power, by number of periods $T$ and $\phi=m\rho/(1-\rho)$.}
\label{tab:kgrid}
\small
\begin{tabular}{@{}lccccc@{}}
\toprule
$T$ & $\phi=0.1$ & $0.5$ & $2$ & $10$ & TWFE\\
\midrule
4 & 0.206 & 0.158 & 0.134 & 0.125 & 0.122\\
5 & 0.204 & 0.156 & 0.133 & 0.126 & 0.123\\
6 & 0.201 & 0.152 & 0.132 & 0.125 & 0.123\\
8 & 0.193 & 0.147 & 0.130 & 0.124 & 0.123\\
10 & 0.186 & 0.143 & 0.128 & 0.124 & 0.123\\
\bottomrule
\end{tabular}
\end{table}

\begin{table}[h]
\centering
\begin{threeparttable}
\caption{Stress tests with $G=24$, $T=5$. Variances relative to IT in the balanced complete wedge under the same scenario. The ETI-optimal design is computed with at most 6 clusters starting per period.}
\label{tab:stress}
\small
\begin{tabular}{@{}lcccccc@{}}
\toprule
 & \multicolumn{2}{c}{Complete wedge} & \multicolumn{1}{c}{8 wait-listed} & \multicolumn{3}{c}{ETI-optimal hybrid}\\
\cmidrule(lr){2-3}\cmidrule(lr){4-4}\cmidrule(lr){5-7}
Scenario & IT $\kappa^*$ & ETI cost & IT $\kappa^*$ (var.) & design & ETI cost & IT $\kappa^*$\\
\midrule
Exchangeable, $\phi=2$ & 0.133 & 2.27 & 0.214 (0.93) & 6/6/1/0+11 & 1.23 & 0.38\\
Exchangeable, $\phi=10$ & 0.126 & 2.51 & 0.199 (0.96) & 6/5/2/0+11 & 1.34 & 0.31\\
Block exch., CAC $0.8$ & 0.140 & 2.12 & 0.226 (0.91) & 6/5/1/0+12 & 1.15 & 0.44\\
Block exch., CAC $0.5$ & 0.156 & 1.87 & 0.254 (0.87) & 6/6/0/0+12 & 1.02 & 0.55\\
Exponential decay, $r=0.8$ & 0.131 & 2.47 & 0.193 (1.00) & 6/5/1/0+12 & 1.41 & 0.31\\
Exponential decay, $r=0.5$ & 0.152 & 1.85 & 0.241 (0.88) & 6/5/0/0+13 & 1.03 & 0.49\\
Binary, logistic GLMM\tnote{a} & 0.128 & 2.28 & 0.206 (0.92) & 6/5/1/0+12 & 1.23 & 0.41\\
HH working model, CAC $0.5$\tnote{b} & 0.133 & 2.09 & 0.214 (0.90) & 6/5/1/0+12 & 1.14 & 0.41\\
\bottomrule
\end{tabular}
\begin{tablenotes}\footnotesize
\item[a] Linearised logistic mixed model, 50 individuals per cell, baseline prevalence 0.2, log odds ratio 0.4, cluster variance 0.15.
\item[b] Analyst fits the Hussey--Hughes model; the truth is block-exchangeable with CAC 0.5.
\item Designs list the clusters starting in periods 2/3/4/5 plus those wait-listed; $\phi=2$ unless stated.
\end{tablenotes}
\end{threeparttable}
\end{table}

\begin{table}[h]
\centering
\caption{Smallest ETI cost among designs with at most $k$ wait-listed clusters and at most 6 starting per period ($G=24$, $T=5$); the optimal design (clusters starting in periods 2/3/4/5 plus wait-listed) in brackets.}
\label{tab:frontier}
\footnotesize
\setlength{\tabcolsep}{4pt}
\begin{tabular}{@{}lcccccc@{}}
\toprule
Scenario & $k=0$ & $k\le4$ & $k\le6$ & $k\le8$ & $k\le12$ & PB 12/12\\
\midrule
Exchangeable, $\phi=2$ & 2.27 & 1.52 & 1.38 & 1.29 & 1.23 & 1.46\\
 & & [6/6/2/6+4] & [6/6/0/6+6] & [6/6/1/3+8] & [6/6/1/0+11] & \\
Exchangeable, $\phi=10$ & 2.51 & 1.65 & 1.50 & 1.40 & 1.34 & 1.76\\
Block exch., CAC $0.5$ & 1.87 & 1.28 & 1.16 & 1.09 & 1.02 & 1.04\\
Binary, logistic GLMM & 2.28 & 1.54 & 1.39 & 1.30 & 1.23 & 1.44\\
\bottomrule
\end{tabular}
\end{table}

\begin{table}[h]
\centering
\begin{threeparttable}
\caption{Coverage of TATE by nominal 95\% intervals in simulated balanced complete wedges ($T=5$, $\phi=2$, linear ramp; 20{,}000 replications per row).}
\label{tab:smallG}
\small
\begin{tabular}{@{}lcccccc@{}}
\toprule
 & & \multicolumn{3}{c}{TWFE-IT} & GLS-IT & TWFE-ETI\\
\cmidrule(lr){3-5}\cmidrule(lr){6-6}\cmidrule(lr){7-7}
$\kappa$ & $G$ & true SD, $z$ & CR1, $t_{G-1}$ & CV3, $t_{G-1}$ & model SE & CV3, $t_{G-1}$\\
\midrule
0 & 8 & 0.950 & 0.935 & 0.963 & 0.951 & 0.959\\
0.15 & 8 & 0.880 & 0.874 & 0.919 & 0.890 & 0.959\\
0.30 & 8 & 0.675 & 0.703 & 0.790 & 0.700 & 0.959\\
0.30 & 16 & 0.682 & 0.688 & 0.734 & 0.700 & 0.955\\
0.30 & 24 & 0.673 & 0.684 & 0.712 & 0.697 & 0.956\\
\bottomrule
\end{tabular}
\begin{tablenotes}\footnotesize
\item CV3 standard error relative to the true SD: 1.05 ($G=8$), 1.01 ($G=16$), 1.02 ($G=24$). The data-generating model includes a linear period trend.
\end{tablenotes}
\end{threeparttable}
\end{table}

\begin{table}[h]
\centering
\begin{threeparttable}
\caption{Digital adherence trial \citep{cattamanchi2021digital}: fits, diagnostic and counterfactual redesign. Outcome: treatment success; odds ratios unless stated.}
\label{tab:uganda}
\footnotesize
\setlength{\tabcolsep}{4pt}
\begin{tabular}{@{}ll@{}}
\toprule
\multicolumn{2}{@{}l}{\emph{Fits (transition months excluded; 1{,}913 patients, 18 facilities)}}\\
IT, model-based 95\% CI & 1.04 (0.68, 1.58)\\
IT, CV3 with $t_{17}$ & (0.55, 1.97); CV3 SE / model SE $=1.41$\\
ETI, TATE, model-based 95\% CI & 1.01 (0.53, 1.91)\\
ETI, TATE, CV3 with $t_{17}$ & (0.47, 2.16); CV3 SE / model SE $=1.11$\\
ETI exposure effects, $e=1,\dots,6$ & 1.03, 1.06, 0.93, 1.39, 1.21, 0.63\\
\quad log-scale SE; patients & 0.24, 0.28, 0.34, 0.39, 0.50, 0.62; 270, 228, 156, 129, 73, 35\\
Constant-effect test; slope of $\hat\theta_e$ on $e$ & $\chi^2_5=3.93$, $p=0.56$; $0.010$ (SE $0.079$)\\
\addlinespace
\multicolumn{2}{@{}l}{\emph{Diagnostic (log-odds scale, linearised at the IT fit)}}\\
$\hat\phi$; patients per facility-month; RE SD & 0.60; 15.4; 0.46\\
IT exposure weights & 0.77, 0.39, 0.08, $-0.06$, $-0.11$, $-0.07$\\
$\Gamma$; $\kstar$ & 1.91; 0.122\\
$\kstar$ of the design with equal cells: $\hat\phi$, $\phi=2$, TWFE & 0.122, 0.110, 0.103\\
$\delta_{80}$; tolerated spread $\kstar\delta_{80}$ & 0.605; 0.074\\
$(W-u)^\top\hat\theta$ (SE) & 0.021 (0.237)\\
Spread of $\hat\theta$: raw; estimation error; debiased & 0.250; 0.259; 0\\
Power of constant-effect test, spread $=1,2,3,5$ radii\tnote{a} & 0.08, 0.18, 0.37, 0.85\\
\quad IT bias at that spread, in IT SEs & 0.65, 1.30, 1.96, 3.26\\
\addlinespace
\multicolumn{2}{@{}l}{\emph{Counterfactual (equal cells, $\phi=0.60$, at most 3 switching per month)}\tnote{b}}\\
Wait-listed: 0 / 2 / 4 / 6 / 8 & \\
\quad ETI variance / IT variance of the trial's design & 2.54 / 1.73 / 1.41 / 1.25 / 1.18\\
\quad IT variance / IT variance of the trial's design & 1.00 / 0.90 / 0.88 / 0.90 / 0.98\\
\quad IT radius $\kstar$ & 0.128 / 0.162 / 0.198 / 0.254 / 0.366\\
\quad Facilities starting in months 1--6 & 3,3,3,3,3,3 / 3,3,3,1,3,3 / 3,3,3,0,2,3 / 3,3,3,1,0,2 / 3,3,3,1,0,0\\
\bottomrule
\end{tabular}
\begin{tablenotes}\footnotesize
\item Logistic mixed model with random facility intercept, categorical month, and sex, age, HIV status, disease class, retreatment and facility type; maximum likelihood, 30-point Gauss--Hermite quadrature.
\item[a] Least-favourable direction $\theta-\tau\one\propto W-u$; Wald test with the ETI covariance; 5\% level.
\item[b] Transition month counted as exposure 1 ($T=8$, $L=7$), as in the trial's prespecified sensitivity analysis. Outcomes under these designs were not observed.
\end{tablenotes}
\end{threeparttable}
\end{table}

\clearpage
\section{Reproducibility map}\label{app:repro}

\begin{table}[h]
\centering
\small
\begin{tabular}{@{}>{\raggedright\arraybackslash}p{0.45\textwidth}>{\raggedright\arraybackslash}p{0.5\textwidth}@{}}
\toprule
Result & Code\\
\midrule
\Cref{lem:balance,lem:wb}, \cref{thm:twfe,thm:gls,thm:hybrid}, \cref{prop:ramp,prop:t3,prop:parallel}; closed-form values in the text & \texttt{verify\_closed\_}\allowbreak\texttt{forms\_exact.py}\\
\Cref{thm:hybridgls} & \texttt{proof\_tate\_}\allowbreak\texttt{hybrid.py}, \texttt{proof\_tate\_}\allowbreak\texttt{hybrid\_check.py}\\
\Cref{tab:levers}, \cref{nr:realloc} ($G=24$) & \texttt{care\_sw\_pilot.py} \S\S2--4, 6; \texttt{gh\_comparator.py}\\
\Cref{nr:realloc} (grids), \cref{tab:kgrid} & \texttt{ceiling.py}, \texttt{replicate.py}\\
\Cref{nr:frontier}, \cref{fig:etcost}, \cref{tab:frontier} & \texttt{frontier\_waitlist.py}\\
\Cref{nr:etiopt}, \cref{tab:stress}; constant-effect comparator & \texttt{stress.py}; \texttt{gh\_comparator.py}\\
\Cref{tab:curves} & \texttt{curves.py}\\
\Cref{tab:honest}, \cref{prop:crossover} values & \texttt{care\_sw\_pilot.py} \S\S5, 8\\
\Cref{tab:smallG}, Monte Carlo checks & \texttt{care\_sw\_pilot.py} \S9; \texttt{mc\_check.py}\\
\Cref{sec:example}, \cref{tab:haines_fits,tab:haines}, \cref{fig:haines} & \texttt{haines\_trial1\_analysis.py} (data: S2 of \citealp{haines2017impact}); \texttt{haines\_cti\_}\allowbreak\texttt{aic\_check.py}\\
\Cref{sec:uganda}, \cref{tab:twotrials,tab:uganda} & \texttt{uganda\_}\allowbreak\texttt{reanalysis.py}, \texttt{uganda\_}\allowbreak\texttt{reanalysis\_extra.py} (data: S1 Data of \citealp{cattamanchi2021digital})\\
\bottomrule
\end{tabular}
\end{table}

\end{document}